\documentclass[prb, aps,
twocolumn,
 showpacs,amsmath,amssymb]{revtex4}
\usepackage{epsf}
\usepackage{graphicx}
\def \be {\begin{equation}}
\def \ee {\end{equation}}

\def \bea {\begin{align}}
\def \eea {\end{align}}
\def \p {\partial}
\def \BEA {\begin{eqnarray}}
\def \EEA {\end{eqnarray}}

\def \f {\varphi}

\begin{document}

\title{Conductivity  of   suspended graphene at the Dirac point
}
\author{I. V. Gornyi$^{1,2}$}
\author{ V. Yu. Kachorovskii$^{1,2,3}$}
\author{A. D. Mirlin$^{1,3,4}$}
\affiliation{$^{1}$Institut f\"ur Nanotechnologie,  Karlsruhe Institute of Technology,
76021 Karlsruhe, Germany
\\
$^{2}$    A.F.Ioffe Physico-Technical Institute,
194021 St.~Petersburg, Russia
\\$^3$ Institut f\"ur Theorie der kondensierten Materie,  Karlsruhe Institute of
Technology, 76128 Karlsruhe, Germany
\\$^{4}$ Petersburg Nuclear Physics Institute, 188300, St.Petersburg, Russia
}

\date{\today}

\begin{abstract}
We study transport properties of clean suspended graphene
at the Dirac point. In the absence of the electron-electron interaction, the
main contribution to resistivity comes from interaction with flexural
(out-of-plane deformation) phonons. We find that the  phonon-limited
conductivity
scales with the temperature as $T^{-\eta}, $ where $\eta$ is the
critical exponent (equal to $\approx 0.7$ according to numerical
studies) describing renormalization of the flexural phonon correlation functions
due to anharmonic coupling   with the in-plane phonons. The electron-electron
interaction induces an additional scattering mechanism and also affects the
electron-phonon scattering by screening the deformation potential.
 We demonstrate that the combined effect of both interactions results in a
conductivity that can be expressed as a dimensionless function  of two
temperature-dependent dimensionless
 constants,  $G[T]$ and $G_e[T],$  which characterize the strength of
electron-phonon and electron-electron interactions, respectively.
We also discuss the behavior of conductivity away from the
Dirac point as well as the role of the impurity potential and compare our
predictions with available experimental data.

\end{abstract}
\maketitle

\section{Introduction}
\label{s1}

The discovery of graphene, a single monolayer of graphite, \cite{Geim,Geim1,Kim}
has initiated a remarkably intensive study  of electronic
properties of graphene structures (for review, see
Refs.~\onlinecite{geim07,graphene-review}).
 This interest has both  fundamental reasons and application-related
motivations.  From the fundamental point of view, the interest to graphene is
largely motivated by the quasirelativistic character of its spectrum: charge
carriers in graphene are two-dimensional (2D) massless Dirac fermions. This
leads to a variety of remarkable phenomena related to inherent topology of
Dirac fermions as well as to their properties in the presence of various
types of disorder and interactions. Further, the Dirac character of spectrum
makes graphene a unique example of a system where essentially quantum
phenomena such as the quantum Hall effect can be observed up to the room
temperature. \cite{novoselov07}
From the prospective of applications, the technological breakthrough in
fabrication of flat single-layer  2D systems opens a wide avenue for creation
of ultimately thin 2D nanostructures, thus being in the mainstream  of  the
general tendency to miniaturization  of   electronic devices. Moreover,
suspended graphene samples demonstrate the room-temperature mobility
as high as  $1.2\times10^5$ cm$^2$/Vs, which is higher than for
conventional semiconductor 2D structures. Therefore,
high-quality suspended graphene flakes with the size of the order of 1 $\mu$m
may show ballistic transport up to the room temperature. \cite{Bolotin0, Du,
Bolotin,Meyer,McEuen,Miao07,Danneau08,tension,Lau}  It is
widely believed that, in combination with carbon nanotubes, graphene may form a
basis for the future carbon electronics. Hence  investigation of transport
properties of graphene is a highly topical problem.

At low temperatures, the resistivity of graphene is dominated by scattering off
impurities. Away from the Dirac point, the dependence of graphene
conductivity on electron concentration depends strongly
on the nature of scatterers. \cite{impurity} The experimentally observed
(approximately linear) dependence in most of the samples may
be explained by strong impurities creating resonances
near the Dirac point (``midgap states''), \cite{impurity,Stauber07} yielding
$\sigma\propto n \ln^2 n$, or, alternatively, by Coulomb impurities and/or
ripples, leading to $\sigma \propto n.$
\cite{impurity,Ando06,Nomura07,Morpurgo06} The
dominant type (or types) of disorder and the corresponding disorder strength
depend, of course, on technology of the sample preparation.

A hallmark of Dirac nature of carriers in graphene
is the minimal conductivity $\sim e^2/h$ at the Dirac point. \cite{Geim1,Kim}
Remarkably, it was found experimentally that the minimal conductivity stays
almost unchanged up to a very low temperature ($\sim 30$\:mK, i.e., three order of
magnitudes below the  impurity-induced transport relaxation rate).\cite{Tan07} This can be
explained by
``protection'' of disordered Dirac fermions from quantum localization in the
absence of intervalley scattering \cite{ostrovsky07} or in the case
of a chiral-symmetric disorder.\cite{impurity,ostrovsky10}

At higher temperatures, the graphene resistivity is expected to be
determined by electron-phonon and electron-electron interactions.
Manifestations of both interactions in transport properties of graphene have
been studied in the literature; see
Refs.~\onlinecite{Kashuba, Fritz,aleiner,Sachdev,aip,fluid,Foster,Ryzhii,Schuett-ee,nguyen,svintsov}
for discussion of the role of electron-electron collisions  and
Refs.~\onlinecite{Mahan,Suzuura,Hw,Manez,CaKim,Fasolino,aleiner-basko,basko,Oppen-short,Oppen-Scr,
Oppen-long,Vozmediano,Ochoa,kat2,kat1,San-Jose, Ochoa1} for  discussion of
electron-phonon scattering. In this connection, two important features
distinguishing graphene from conventional 2D semiconductor systems should be
emphasized. First,  at the Dirac point of graphene, the electron-electron
scattering leads to velocity relaxation (though total momentum is
conserved just as in the conventional case) and, therefore,  gives a
contribution to resistivity. \cite{Kashuba,Fritz,aleiner,Sachdev,Ryzhii,Schuett-ee}
Second, a suspended flake of graphene is a crystalline membrane, which
implies existence  of
specific type of  the phonon modes, so-called flexural phonons. \cite{Nelson}

Apart from their role as one of the most important scattering mechanisms for
electrons, the flexural phonons are very interesting from the point of view of
mechanical properties and thermodynamic stability \cite{mermin,landau} of graphene membrane.
The
out-of-plane fluctuations represent a particularly soft mode
($\omega\sim q^2$ dispersion when anharmonicity is neglected versus $\omega \sim q$
for conventional phonon modes), so that they might
be expected to be very efficient in inducing
strong thermal out-of-plane fluctuations and thus driving the membrane into the
so-called crumpled phase. This question was intensively discussed in the
literature two decades ago
\cite{Nelson0,NelsonCrumpling,Doussal} in connection
with biological membranes, polymerized layers, and some inorganic surfaces (see
also the review in Ref.~\onlinecite{Nelson}  as well as more recent papers,
Refs.~\onlinecite{Xeng} and \onlinecite{eta1}).
It was found  that anharmonic coupling of in-plane and out-of-plane phonons
stabilizes the membrane for sufficiently low temperatures $T$, so
that the membrane is in the flat phase at relatively low $T$ and undergoes the
crumpling transition with increasing $T$. The main dimensionless parameter
characterizing the state of the membrane is the ratio of the bending rigidity
$\kappa$ to the temperature. For graphene, this ratio for room temperature is
quite large, $\kappa/T \simeq 30$. This reflects a remarkable rigidity of
graphene and implies that graphene remains in the flat phase up to the temperatures
several times higher than the room temperature.

In this paper, we explore  transport properties   of  clean suspended
graphene, with a particular focus on the case of zero chemical potential (Dirac
point). We show that, despite a high bending rigidity, flexural phonons play
central role in determining the graphene resistivity in a broad range of
temperatures. We also demonstrate that the anharmonicity crucially affects the
magnitude and the temperature dependence of the resistivity.

The structure of the paper is as follows. In Sec.~\ref{s2}, we carry out a
general analysis of
electron-phonon scattering  in a suspended graphene. First, we neglect the electron-electron interaction.
In this case, the main contribution to resistivity comes from scattering by
deformation  flexural phonons, while  other  types of electron-phonon
interaction can be neglected for realistic values of temperatures.
We predict  a power-law  dependence of the transport scattering rate on
the energy and temperature,
$1/\tau_{\rm{tr}} \propto |\epsilon|^{2\eta-1} T^{2-\eta},$ where  energy,
$\epsilon,$ is counted from the Dirac point and  $\eta \approx 0.7 $
is the critical exponent describing renormalization  of
 the flexural-phonon  correlation functions due to anharmonic coupling   with
the  in-plane phonons. As a result, the  phonon-limited conductivity at the
Dirac point scales with the temperature as $T^{-\eta}.$

In Sec.~\ref{s3}, we take
the electron-electron interaction into consideration and demonstrate  that
its effect is twofold: (i) it screens the deformation potential
 and  (ii) it induces an additional scattering channel.
We show that  both these effects may be taken into account on equal footing
by introducing two dimensionless temperature-dependent constants,  $G(T)$ and
$G_e(T)$, which characterize the strength of electron-phonon and
electron-electron interaction, respectively. We find the conductivity and
demonstrate that it can be expressed as a dimensionless function of $G$ and
$G_e$ (i.e., all the temperature dependence can be absorbed in these two
parameters).
Different transport regimes correspond thus to different parts of $(G,G_e)$
plane (see Fig.~\ref{Fig2}).
At low temperatures electron-electron collisions are more intensive ($G_e >G$),
while  at relatively high temperatures,
$G$
becomes larger than $G_e$ and  flexural phonons dominate. When the impurity
scattering  is also taken into account, the temperature dependence of
conductivity at the Dirac point becomes strongly non-monotonous at low $T.$

In Sec.~\ref{s4}, the behavior of conductivity away from the Dirac point
(non-zero chemical potential $\mu$) is discussed. We show that at $\mu \ne 0$
the interplay of electron-electron and electron-phonon interactions leads to a
highly non-trivial temperature dependence of conductivity containing several
regions with different power-law behavior.  Taking into account impurity
scattering makes the whole picture even more complex. We analyze how  the
temperature dependence of conductivity evolves with
increasing $\mu$ from a non-monotonous (with a maximum in the intermediate
temperature range) at low $\mu$ to monotonously decaying at sufficiently high
$\mu$.

In Sec.~\ref{s5} we  compare our findings
with available experimental data and find  a very good qualitative agreement between theory and experiment.

Section \ref{s6} contains a summary of obtained results. We also discuss there open questions and
prospective directions for future research.

Technical details of the calculation of scattering rates and the hydrodynamic approach are relegated
to Appendixes \ref{a1} and \ref{a2}, respectively.

\section{Flexural phonons}
\label{s2}

\subsection{Elasticity of graphene}
\label{s2.1}

 We start with a discussion of elastic properties of graphene.
  The
energy of the in-plane and out-of-plane elastic deformations of graphene
 is given by conventional expression known from membrane physics: \cite{Nelson,graphene-review}
 \be
    E = \frac{1}{2} \int d{\bf r}\left [{\rho} (\dot{\bf u}^2+\dot h^2)+ \varkappa (\Delta h)^2 + 2 \mu
    u^2_{ij}+ \lambda u^2_{kk}\right].
\label{energy}
\ee
 Here  ${\bf u}({\bf r})$  and $h({\bf r})$  are the in-plane and  out-of-plane distortions,
  \be u_{ij}=\frac{1}{2}[\partial_i u_j + \partial_j u_i + (\partial_i h) (\partial_j h)] \label{distortions}\ee
  is
  the strain tensor,   $\rho \simeq 7.6 \times 10^{-7}$ kg/m$^2$ is the mass density of graphene,
 $\lambda \simeq 3$ eV/\AA $^2$ and $\mu \simeq 3$ eV/\AA $^2$ are in-plane
elastic constants, and $\varkappa \simeq 1$ eV is the bending rigidity.

As follows from  Eq. \eqref{energy}, there are three types of acoustic phonons
in graphene:
longitudinal ($\parallel$) and transverse ($\perp$) in-plane  modes,  and  out-of-plane  flexural mode.  The mode frequencies read
\be \label{omega}
\omega_{\parallel \mathbf{q}}=s_{\parallel}q\:, \qquad \omega_{\perp
\mathbf{q}}=s_{\perp}q\:, \qquad \omega_{ \mathbf{q}}= D q^2\:,
\ee
where $s_{\parallel}=\left[\left(2\mu +\lambda\right)/\rho\right]^{1/2}\simeq 2\times 10^6_{}\, {\rm cm}/{\rm s},$  $s_{\perp}=\left(\mu/\rho\right)^{1/2} \simeq 1.3\times 10^6_{}\, {\rm cm}/{\rm s} $, and $D=\sqrt{\varkappa/\rho}\simeq 0.46\times 10^{-2}~ {\rm cm}^2/{\rm s}.$\

Below, we assume that $qa\ll 1$ (here $a$ is the bond length of the honeycomb  graphene lattice), thus neglecting intervalley transitions. In this approximation, we can focus on the study of  the vicinity of one of two equivalent Dirac points, writing  the electron Hamiltonian as
\be
H=\hbar v \boldsymbol{\sigma} \mathbf{k},
\label{ham-e}
\ee
where $v \simeq 1 \times 10^8~ {\rm cm}/{\rm s} $ is the Fermi velocity,
$\boldsymbol{\sigma}$ is vector consisting of Pauli matrices acting in the
sublattice space, and wave vector $\mathbf{k}$  is measured from the Dirac point.
 Eigenfunctions of this Hamiltonian and    corresponding eigenenergies are given
by
\be
\psi_{\mathbf k \alpha}=e^{i \mathbf k \mathbf r}\vert \chi_{\mathbf k}^{
\alpha} \rangle, \qquad \epsilon_\alpha(\mathbf k)= \alpha\hbar v k,
\ee
where $\alpha=\pm,$ and
\be
\vert \chi_{\mathbf k}^{ \alpha} \rangle =\frac{1}{\sqrt{2}}\left ( \begin{array}{c}
                                                  e^{-i\f_\mathbf k/2} \\
                                                   \alpha e^{i\f_\mathbf k/2}
                                                \end{array}\right),
\ee
with $\f_\mathbf k$ denoting the polar angle of the momentum $\mathbf k$.

The electron-phonon interaction Hamiltonian reads
\be
H_{e,ph}=\int d \mathbf r \Psi^\dagger  V_{e,ph} \Psi,
\label{ham-e-ph}
\ee
where $\Psi({\bf r})$ is the electron field operator,
\be
\Psi^\dagger=\sum_\mathbf k   a^{ \dagger}_{\mathbf k\alpha}\psi_{\mathbf k
\alpha}, \qquad
\Psi=\sum_\mathbf k   a_{\mathbf k\alpha}\psi_{\mathbf k \alpha}^*
\label{Psi}
\ee
and $V_{e,ph}$ can be separated into the deformation potential (diagonal in
sublattice space) and the effective gauge field (off-diagonal in the sublattice
space), \cite{Suzuura}
\be
V_{e,ph} = V + V_\mathbf A = g_1  u_{ii} + g_2\boldsymbol{\sigma}\mathbf A.
\label{V-e-ph}
\ee
Here  $g_1 \simeq 30~ {\rm eV} $ is the  bare (unscreened) deformation coupling
constant, and $g_2 \simeq 1.5~ {\rm eV} $ corresponds to coupling to  the
phonons via the effective gauge field $\mathbf A =(A_x,A_y)$ with the components
\be
A_x= 2u_{xy}\:, \qquad A_y= u_{xx}-u_{yy}\:.
\label{gauge}
\ee
Equations (\ref{V-e-ph}) and (\ref{gauge}) represent contributions of leading
order (in gradients of the distortion fields) to the phonon-induced scalar and
vector potential, respectively. Higher-order terms \cite{graphene-review} would only give
small corrections and are not considered below.

\subsection{Quasielastic scattering by flexural phonons }
\label{s2.2}

Since $g_1 \gg g_2$,  we will first consider the  deformation part of the
electron-phonon  potential that is expected to give a dominant contribution to
the scattering rate. Contribution of the gauge field will be discussed later;
we will see that it is indeed much smaller.
Further, we will focus on the contribution of the flexural phonons to the
deformation potential
\be
V =g_1 (\nabla h)^2/2 \,,
\label{V}
\ee
which is larger than that of longitudinal phonons because of softer dispersion
of flexural modes.

The transverse displacement field $h(\mathbf r)$ can be written as
\be
h(\mathbf r)=\sum_{\mathbf q} {\sqrt\frac{\hbar}{2\rho\omega_\mathbf q S
}}(b_\mathbf q +b^\dag_{-\mathbf q} ) e^{i \mathbf q\mathbf r},
\label{h}
\ee
where $S$ is the sample area. Since $V$ is quadratic with respect to $h,$
electron in any scattering act emits (absorbs) two phonons with the wave vectors
$\mathbf q_1$ and $\mathbf q_2,$ the total transferred wave vector being
$\mathbf Q=\pm \mathbf q_1 \pm  \mathbf q_2.$  The transport scattering rate,
$1/\tau_{\rm{tr}}(\epsilon),$
can be presented as integral over $d^2\mathbf q_1 d^2 \mathbf q_2.$
 The main contribution to the integral comes from the region in $(\mathbf q_1, \mathbf q_2)$ space,
 where one of the momenta, say $q_2,$ is much smaller than the other, and  the
integral is logarithmically  divergent $ \propto \int dq_2/q_2.$ The upper limit
of the integral is given by  $\epsilon/\hbar v,$
 while the lower limit is given by inverse sample size $1/L$ provided that
nonlinear interaction of flexural phonons with in-plane phonons
 is neglected. In fact, such anharmonicity provides an infrared cutoff for the
logarithmic divergency due to screening of flexural phonons by in-plane ones;\cite{Doussal}
we will first neglect it and include later into
consideration.
Since $q_2 \ll q_1,$ we find that total transferred momentum is given by $\mathbf
q_1$ and total energy gained (or lost) by electron is given by $\hbar\omega_{\mathbf q_1}=  D q_1^2 \propto
\hbar D (\epsilon/\hbar v)^2.$   This energy is much smaller than $T$ [see
Eqs.~ \eqref{q*}, \eqref{ineq} below and discussion after Eq.~\eqref{ineq}], so that phonon Planck
numbers are large  and     one can replace  $b_{\mathbf q}$ in Eq.~\eqref{h}
with $\sqrt{T/\hbar \omega_\mathbf q} \exp(-i\f_\mathbf q),$ where $\f_\mathbf
q$ are  random phases (with the correlation function $\langle \exp[i(\f_\mathbf
q-\f_{\mathbf q^\prime})]\rangle=\delta_{\mathbf q,\mathbf q^\prime}$), over
which the final expression for the scattering rate should be averaged.    The
inequalities   $\hbar\omega_{\mathbf q_1} \ll T,\hbar\omega_{\mathbf q_2} \ll T$ also imply that the phonon potential is
quasistatic.  Thus, to the leading approximation, one can  assume  that
$h(\mathbf r)$ is a static field,
\be
h(\mathbf r)=\sum_\mathbf q\sqrt{\frac{2T}{\varkappa q^4 S}} \cos(\mathbf q \mathbf r +\f_\mathbf q).
\label{static-field}
\ee
As seen from Eq.~\eqref{static-field}, the  rms thermal fluctuation  of  the out-of-plane amplitude,
\be
\sqrt {\langle h^2(\mathbf r) \rangle} \propto  \sqrt{\frac{T}{\kappa}  \int \frac{d^2 \mathbf q}{q^4}} \propto \sqrt{\frac{T}{\kappa}} L ,
\label{rms}
\ee
 is proportional to the system size $L$ (when anharmonicity is neglected) and the ratio $\sqrt {\langle h^2(\mathbf r) \rangle}/L$
 is controlled by  the dimensionless parameter $T/\kappa.$

Evaluating the transport rate for the scattering on the quasistatic random potential given by Eqs.~\eqref{V} and \eqref{static-field}
(see Appendix \ref{a1-1}), we find
\be
\frac{1}{\tau_{\rm{tr}}(\epsilon)}= \frac{ 2g^2 T^2}{\pi \hbar |\epsilon|} \ln \left(
\frac{|\epsilon|}{v L} \right)\,,
\label{tau-tr-final}
\ee
for a particle with energy $\epsilon$  counted from the Dirac point.
Here
\be
g=\frac{g_1}{\sqrt{32}\varkappa} \simeq 5.3
\label{g}
\ee
is the dimensionless coupling constant.  We see that the coupling is quite
strong and is additionally enhanced by a divergent logarithm.

As has been already mentioned, the above analysis in fact overestimates the
thermal fluctuations related to flexural phonons. The divergent logarithm in
Eq.~\eqref{tau-tr-final} appeared because we studied flexural phonons in the harmonic
approximation. We are now going to take into account the anharmonic
phonon-phonon interaction.  As known
from the membrane theory, \cite{Doussal,Nelson}  anharmonic coupling of the
flexural phonons with the in-plane ones leads to screening of the phonon Green
function
\be G_q= \langle h_\mathbf q h_{\mathbf q}^\dagger \rangle, \label{Gh}\ee
where $h_\mathbf q =b_\mathbf q\sqrt{\hbar/ 2\rho \omega_\mathbf q}$. Specifically,
while in the harmonic approximation, $G_q \propto T/\varkappa q^4 $ [see
Eqs.~\eqref{static-field} and \eqref{rms}], the anharmonic coupling   suppresses
flexural oscillations by modifying the power-law behavior at
large scales, i.e., for $q$ smaller than certain $q_c.$ The value of $q_c$ for
graphene can be estimated by accounting of interaction between out-plane  and
in-plane modes in the framework of the perturbation theory: \cite{Fasolino,Ochoa1}
\be
q_c=\frac{\sqrt{T\Delta_c}}{\hbar v}, \qquad \Delta_c=\frac{3\mu v^2(\mu
+\lambda)\hbar^2}{4\pi\varkappa^2 (2\mu+\lambda)} \simeq 18.7 ~{\rm eV}.
\label{qc}
\ee
Let us compare $q_c$ to another important momentum scale $q_*$ that
is determined by the condition $\hbar \omega_{q_*}= \hbar D q_*^2   = T$,
\be
q_*=\frac{\sqrt{T\Delta_*}}{\hbar v}, \qquad \Delta_*=  \frac {\hbar v^2}{D}
\simeq
1.25 \cdot 10^3 ~{\rm eV}.
\label{q*}
\ee
For $q_1,q_2 \ll  q_*$ the electron-phonon scattering  is quasielastic.
Using above estimates for $\Delta_c$ and $\Delta_*$, we find that
\be
q_T \ll q_c \ll q_*,
\label{ineq}
\ee
where $q_T =T/\hbar v$.  In the Dirac point, the   characteristic momentum
transferred in a scattering act  is of the order of $q_T$ and therefore is
small compared to $q_c$.  Hence the anharmoinic interaction of flexural
phonons with in-plane phonons should be
taken into account. Further, Eq.~\eqref{ineq} ensures that
relevant momenta are small compared to $q_*, $ so that the quasielastic
approximation used above is justified. It is worth noting that this approximation  also applies away from the Dirac point because
inequality $\epsilon/\hbar v\ll q_*$ ($\epsilon \ll \sqrt {T \Delta_*}$) is  typically satisfied for relevant energies  and not too small $T.$

Interaction between flexural  and  in-plane phonons leads to a
power-law renormalization of the bending rigidity: \cite{Doussal,Xeng, Nelson}
\be
\varkappa \to \varkappa(q) \sim \varkappa \left(\frac{q_c}{q}\right)^\eta,~~\text{for}~~ q \ll
q_c,
\label{bending}
\ee
and  the phonon Green function takes a form
\be
G_q = Z \frac {T}{\varkappa q^4} \left( \frac{q}{q_c} \right)^\eta\,, \qquad
q \ll q_c.
\label{G-eta}
\ee
Here $\eta$ is a critical index and $Z \sim 1$.  On
the analytical level, the value of $\eta$ was found in the limit of
large spatial dimensionality; \cite{Aronovitz89} extrapolation of
this result to the situation of interest (2D membrane embedded in a 3D space)
yields $\eta=2/3$.  Some modified versions of the large-dimensionality
approximation have been developed, such as the self-consistent screening
approximation (SCSA) \cite{Doussal,Xeng, Nelson} and the ``non-perturbative
renormalization group''; \cite{eta1} the corresponding results
after extrapolation to the physical dimensionality yield $\eta=0.821$ and
$\eta=0.849$, respectively. Clearly, the extrapolation is not controlled
parametrically; the scattering between the above three values may serve as a
rough estimate of their accuracy. Numerical simulations of the problem
gave values  $\eta=0.60 \pm 0.10$
and $\eta=0.72 \pm 0.04$ (see Ref.~\onlinecite{Gompper91} and Ref.~\onlinecite{Bowick96}, respectively).
We will use the latter value for estimates below. \cite{Zhang93}
As to the numerical prefactor $Z \sim 1$, we did not find  its reliable
value in the literature.
In order to obtain $Z$ theoretically, one should
perform a microscopic modeling of elastic properties of a graphene membrane.
Alternatively, when the anharmonicity regime will be identified in experiment,
one should be able to find $Z$ from a comparison of experimental data with the
theory.
Numerical solution of the SCSA equation  \cite{kat2,kat1}
yielded $Z \approx 3.5$, which is the  only  numerical value  available in the literature. It is worth noting, however, that SCSA, which becomes exact in the limit of
infinite dimensionality, is an uncontrolled approximation for a 2D membrane in a
3D space.   For estimates below,  we   use $Z \simeq 2,$  which allows us to get a  qualitative agreement   with experiment (see, Sec.~\ref{s5}).

Physically, the renormalization-induced increase   of the bending rigidity, Eq.~\eqref{bending},  is a
manifestation of the  tendency of the membrane towards the flat phase, which is
realized when the
bare rigidity is large: $\varkappa/T \gg 1$. In the opposite situation, the
membrane is in the crumpled phase; see Ref.~\onlinecite{Nelson} for
review of the crumpling transition between these two phases. As has been
mentioned above, the ratio $\varkappa/T$ is on the order of 30 for
graphene at room temperature, so that graphene is in the flat phase in the whole
range of temperatures under interest.

Using Eq.~\eqref{G-eta}, we find that  for $q_1,q_2
\ll q_c$ one should  introduce  the cutoff  factors $Z\left( {q_1}/{q_c}
\right)^\eta $ and $Z\left( {q_2}/{q_c} \right)^\eta,$  in the integrand in
Eq.~\eqref{tau-tr}. Assuming
 that $\epsilon \ll \sqrt{T \Delta_*} $, after some algebra [see Appendix \ref{a1-2}]
we get
\BEA \label{tau-tr-appendix}
\frac{1}{\tau_{\rm{tr}}(\epsilon)}&=&\frac{2 g^2T^2}{\pi \hbar|\epsilon|}\\ &\times& \left\{
\begin{array}{ll} \displaystyle{ \ln\left(\frac{|\epsilon|}{\sqrt{T\Delta_c}}
\right) },\ \ & \text{for} ~~|\epsilon| \gg \sqrt{T\Delta_c} \vspace{2mm}  \\ \nonumber
\displaystyle{ C Z^2~\left(
\frac{|\epsilon|}{\sqrt{T\Delta_c}}\right)^{2\eta}},\ \  & \text{for}
~~|\epsilon| \ll \sqrt{T\Delta_c}\,,  \end{array} \right.
\EEA
where $C \simeq 2.26.$ We see that for high energies, $|\epsilon| \gg \sqrt{T\Delta_c},$ the only effect of the anharmonicity is  the  replacement of the infrared cutoff, $q_{\rm{min}} \propto 1/L,$ in Eq. \eqref{tau-tr-final} with the size-independent value $q_c. $ In contrast, at low energies, $|\epsilon| \ll \sqrt{T\Delta_c},$  the scattering rate  is strongly suppressed and goes to zero as $|\epsilon|^{2\eta-1}$ with decreasing  the  energy. The latter case is realized in the Dirac point, where $|\epsilon| \sim T \ll \sqrt{T\Delta_c}.$

Equation~\eqref{tau-tr-appendix}   will be used below for calculation of the  Drude conductivity:
\BEA
\label{Drude}
 \sigma_{\rm{ph}}&=&e^2 N
\int{d\epsilon\,}  {\rho}(\epsilon) \left(-\frac{\p n_F}{\p\epsilon}\right)\,\frac{v^2\tau_{\rm{tr}}(\epsilon)}{2}\\
&=&
\frac{e^2 N}{16\pi\hbar^2}\int_{-\infty}^\infty d \epsilon \frac{|\epsilon| \tau_{\rm{tr}}(\epsilon)}{T \cosh^2[(\epsilon-\mu)/2T]}.
\label{sigma}
\EEA
Here
\be \rho(\epsilon)= \frac{|\epsilon|}{2\pi \hbar^2 v^2}\ee is the density of states in
a single valley, $N=4$ is  the  spin-valley degeneracy of the graphene,
$n_F(\epsilon)=\left\{1+\exp[(\epsilon-\mu)/T]\right\}^{-1}$ is the Fermi-Dirac distribution
function, and $\mu$ is the chemical potential. The cases $\mu=0$ (Dirac point) and $\mu \gtrsim T$ will be discussed in the next two sections.

\section{Conductivity at the Dirac point ($\mu =0$)}
\label{s3}

\subsection{Interplay of electron-electron and electron-phonon scattering rates}
\label{s3.1}

We apply now the above results  to the case of zero
chemical potential (Dirac point).
It is instructive to begin with substituting the expression for
the transport scattering rate with neglected anharmonic coupling,
Eq.~(\ref{tau-tr-final}),  into the Drude formula (\ref{Drude}).
A simple calculation yields
 \begin{equation} \label{Drude1}
 \sigma_{\rm{ph}}=\frac{e^2}{\hbar} \frac{\pi^2 N}{24 g^2 \ln \left( q_T L \right)}
 \qquad \text{(anharmonicity is neglected).}
\end{equation}
Let us compare Eq.~\eqref{Drude1} with the Drude conductivity at the Dirac point limited by electron-electron interaction: \cite{Kashuba}
\be
\sigma_{\rm{ee}}= \frac{e^2}{\hbar} \frac{N\ln^22}{2\pi g_e^2 ({0.69}~ N-0.24)},
\label{Drude-ee}
\ee
where
\be
g_e= \frac{g_e^0}{1+(g_e^0/4)\ln(\Delta/T)}
\label{renorm}
\ee
is a renormalized constant of electron-electron interaction, \cite{gonzalez11}
$g_e^0 =e^2/\hbar \kappa v_F $ is the bare constant, $\Delta$ is the
ultraviolet cutoff (of the order of the bandwidth), and $\kappa$ is the
dielectric constant. In Eq.~\eqref{renorm} it is assumed that $g_e^0 \ll 1$,
so that $g_e$ remains small in course of the renormalization. A more  general
approach presented in Ref.~\onlinecite{aleiner} uses $1/N$ expansion and allows
one to find renormalization group  equations for arbitrary $g_e^0,$ in particular, in the presence of different types of disorder  (see also  Ref.~\onlinecite{aleiner-basko} for analysis of  renormalization  in the presence of   optical  phonons).

    For  graphene suspended in the air,  $g_e^0 $ is estimated theoretically as
$g_e^0 \simeq 2$. (In fact, $g_e$ may be  suppressed by using graphene suspended in the
media with high dielectric constant, for example, in conventional water.) In the absence of disorder and phonons, renormalization   reduces the coupling,  so that $g_e $ becomes small at  sufficiently low
energies (temperatures). Generically, disorder and phonons slow down such decrease. Moreover, relatively strong disorder and/or strong coupling to optical phonons may   even lead to a non-monotonic
dependence of $g_e$ on $T$. \cite{aleiner,aleiner-basko}
We do not discuss this case  here. \cite{footnote}

Below, we will use the renormalized  value of the electron-electron coupling, $g_e,$ as a  parameter of the theory which  can be small,  $g_e \ll 1,$ or on the order of unity.
We will also assume the number of ``flavors'' $N$
to be large, $N \gg 1$.   Another important parameter of the theory,    which can be small or large, is $g_e N$ (both  cases will be discussed below).  Equation~\eqref{Drude-ee} was derived under the assumption $g_e N \ll 1.$            In the limit $ N\gg 1 $,  this equation becomes
\be
\sigma_{\rm{ee}}= \frac{e^2}{h  g_e^2} \frac{\ln^22}{ 0.69 } \,.
\label{Drude-ee-N-large}
\ee
With increasing $g_e,$  $ \sigma_{\rm{ee}}$ decreases and saturates (see discussion at the end of Sec.~\ref{s3.2})  for $g_e N \simeq  1$ at the value\cite{aleiner}
\be
\sigma_{\rm{ee}} \sim \frac{e^2 N^2}{h} .
\label{sigma-ge-large}
\ee
This equation yields an estimate of minimal value of conductivity in the only presence  of electron-electron collisions.

Estimating  now the conductivity limited by electron-phonon and
electron-electron interactions according to Eqs.~(\ref{Drude1}),
(\ref{Drude-ee-N-large}), and (\ref{sigma-ge-large}), we find that
the former is
much smaller,   $\sigma_{\rm{ph}}/\sigma_{\rm{ee}} \sim 10^{-2}$.  Hence,  in the
considered approximation (neglecting the anharmonicity), the electron-phonon
interaction strongly dominates over the electron-electron one.

As discussed in Sec.~\ref{s2.2}, it is important to take into account the
anharmonic coupling between flexural and in-plane phonons that enhances the
bending rigidity and therefore suppresses the electron-phonon scattering rate.
Using the corresponding expression for the scattering rate, Eq.~(\ref{tau-tr-appendix}), and taking into account that $|\epsilon| \sim T \ll \sqrt{T\Delta_c},$ we  obtain from Eq.~\eqref{Drude}
 the phonon-limited Drude conductivity  in  the Dirac
point,
\be
\sigma_{\rm{ph}}\simeq \frac{e^2}{\hbar} \frac{ N C_1}{16 Z^2 C g^2 } \left( \frac{\Delta_c}{T}\right)^\eta.
\label{Drude-final}
\ee
Here, $C_1=\int_0^\infty dx{x^{2-2\eta}}/{\cosh^2(x/2)}\simeq 2.19$.
For room temperature and  $\eta=0.72$
(and approximating $Z$ by 2 as
discussed above), we find $\left(
{\Delta_c}/{T}\right)^\eta \simeq   10^2$, so that $\sigma_{\rm{ph}} \simeq 1.2~
e^2/h$.
Comparing  this estimate  with the one given by Eq.~\eqref{sigma-ge-large}, we  see
that
at room temperature
contribution of the
electron-phonon scattering strongly dominates over the electron-electron one, even for the strong interacting  case, $g_e N \gg 1.$   However, as we will see below, Eq.~\eqref{Drude-final} overestimates the contribution of flexural phonon scattering because it does not take into account screening of the deformation potential. Such a screening leads to  suppression of deformation potential.  As a   result, contributions of electron-phonon and electron-electron   becomes of the same order yielding  a more realistic value  of conductivity  $\sigma_{\rm{ee+ph}} \simeq 10 \div 20 ~
e^2/h$ (see discussion in Sec.~\ref{s5}).

Next, we discuss the competition between the electron-electron and
electron-phonon interactions in the Dirac point  in more detail.  We start with considering a weak
electron-electron interaction $g_e N \ll 1$.
This inequality ensures that the rate of energy relaxation  ${1}/{\tau_E^{\rm{ee}}}$ caused by electron-electron collisions  is much higher  than  the rate of velocity relaxation due to these collisions (in the estimates below we assume $\epsilon \sim T$):  \cite{Schuett-ee}
 \be
 \frac{1}{\tau_E^{\rm{ee}}} \gg \frac{1}{\tau_{\rm{tr}}^{\rm{ee}}}\,.
 \label{ineq1}
 \ee
 Let us first assume that the energy relaxation is also faster than
the phonon-induced transport rate,
  \be
  \frac{1}{\tau_E^{\rm{ee}}} \gg \frac{1}{\tau_{\rm{tr}}^{\rm{ph}}},
  \label{ineq1a}
  \ee
 the relation between  ${1}/{\tau_{\rm{tr}}^{\rm{ee}}}$ and ${1}/{\tau_{\rm{tr}}^{\rm{ph}}}$ being
arbitrary [in the opposite limit, $  {1}/{\tau_E^{\rm{ee}}} \ll
{1}/{\tau_{\rm{tr}}^{\rm{ph}}},$ the electron-electron interaction can be neglected and
conductivity is given by Eq.~\eqref{Drude-final}]. In this case, as a first
step in calculation of conductivity, one should average the inverse scattering
rate over energy for a given direction of velocity: \cite{Schuett-ee}
 \be
 \frac{1}{\tau_{\rm{tr}}^{\rm{ave}}}= \frac{\displaystyle
\left\langle {1}/{\tau_{\rm{tr}}^{\rm{ee}}(\epsilon)}+{1}/{\tau_{\rm{tr}}^{\rm{ph}}(\epsilon)}\right\rangle}{\langle 1 \rangle}\,,
\label{tau-ave}
\ee
where
\be \langle \cdots \rangle= -\int d\epsilon (\cdots) \rho(\epsilon)\p n_F/\p \epsilon .\label{ave}\ee
On the second stage,  one  should
substitute   $\tau_{\rm{tr}}^{\rm{ave}}$  into Eq.~\eqref{Drude}. The result reads
\be  \label{Drude-competition0}
 \sigma_{\rm{ee+ph}} =
 e^2N
\left \langle \frac{v^2\tau_{\rm{tr}}^{\rm{ave}}}{2}\right\rangle =\frac{e^2 N T
\tau_{\rm{tr}}^{\rm{ave}}\ln2}{2\pi\hbar^2} \,.
\ee
Substituting Eq.~\eqref{tau-tr-appendix} into Eq.~\eqref{tau-ave}  and
using Eq.~(76) of Ref.~\onlinecite{Schuett-ee}, we find
\be
\frac{1}{\tau_{\rm{tr}}^{\rm{ave}}}\simeq \frac{T}{\hbar} \left[C_2 Z^2 {g^2 }\left(
\frac{T}{\Delta_c}\right)^\eta + C_3~ {g_{e}^2 N}\right],
\label{tau-ave1}
\ee
where the numerical factors are
$C_2=({2C}/{\pi}) [{\int dx
x^{2\eta}/\cosh^2(x/2)}]/[{\int dx x /\cosh^2(x/2)}] \simeq 2.0$
and $C_3 \simeq 0.989$. Using Eqs.~ \eqref{Drude-competition0} and
\eqref{tau-ave1}, we finally arrive at the following result:
\be
\label{Drude-competition}
 \sigma_{\rm{ee+ph}} =
 \frac{e^2}{\hbar} \frac{N \ln2}{2\pi \left[C_2 Z^2 g^2(T/\Delta_c)^\eta +
C_3 g_{e}^2 N \right]}.
\ee
 Equation \eqref{Drude-competition} may be rewritten as
 \be  \label{Drude-competition1}
 \sigma_{\rm{ee+ph}} =
 \frac{e^2  }{2\pi \hbar}\: N^2 \:\ln2\:  \Sigma(G,G_e),
\ee
where
\be
\Sigma(G,G_e)=\frac{1}{G+G_e}
\label{SigmaGGe}\ee
is a dimensionless function of two dimensionless temperature-dependent coupling
constants, $G= G[T]$ and $G_e=G_e[T]$, defined as
\be
G=  C_2 NZ^2g^2 \left(\frac{T}{\Delta_c}\right)^\eta\,, \qquad G_e= C_3g_e^2N^2
\label{couplings}
\ee
(temperature dependence of $G_e$ is determined by renormalization of $g_e$).
Let us emphasize that all temperature dependence of the conductivity is
absorbed into these two coupling constants. At
low temperatures $G \ll G_e$, so that we have  $\Sigma=1/G_e$ and conductivity
is limited by the electro-electron collisions, while for high temperatures  $G
\gg G_e,$  the electron-phonon scattering dominates and $\Sigma=1/G.$

\subsection{Screening}
\label{s3.2}

Up to now we neglected screening of the electron-phonon interaction by
the electron-electron one. It is known, however, that such a  screening may
reduce deformation coupling constant and thus  the  phonon contribution
to the resistivity. \cite{Hw,Oppen-Scr,Oppen-long}
Quite analogously to Ref.~\onlinecite{Oppen-long} we find that Thomas-Fermi
screening modifies the  electron-phonon coupling:
\be
g \to \frac{g}{ 1+2\pi e^2 N\Pi(Q)/\kappa Q }.
\label{TF4}
\ee
Here $\mathbf Q=\pm \mathbf q_1 \pm \mathbf q_2$ is the total transferred
momentum (see Appendix \ref{a1}) and  $\Pi(Q)=\Pi(\omega,Q)|_{\omega=0}$ is the static
polarization operator. The additional momentum dependence of the
coupling \eqref{TF4} yields the transport scattering rate
\BEA \label{tau-tr-ph}
\frac{1}{\tau_{\rm{tr}}^{\rm{ph}}(\epsilon)} &\sim&
\frac{T}{\hbar} Z^2 g^2
\left(\frac{T}{\Delta_c}\right)^\eta
\left(\frac{|\epsilon|}{T}\right)^{2\eta -1} \\ \nonumber &\times& \left\{
\begin{array}{ll} {\displaystyle {
\left(\frac{\epsilon}{T}\right)^2}\frac{1}{(g_e N)^2}},\ \ & \text{for}~|\epsilon|
\ll g_e N T  \vspace{2mm} \\  1,\ \ & \text{for}~~|\epsilon| \gg g_e N T,
\end{array} \right.
\EEA
calculated in Appendix \ref{a1-3} (hereinafter in this section for the sake of brevity we omit numerical coefficients on the order
of unity). Equation~\eqref{tau-tr-ph} is valid for $g_e N \ll 1.$ In the opposite case, $g_eN \gg 1,$ scattering rate is given by the upper line of Eq.~\eqref{tau-tr-ph} for all relevant energies, $\epsilon \lesssim T.$

Using dimensionless coupling constants, Eq.~\eqref{couplings},
dimensional rate
\be
\gamma=\frac{\hbar N}{T\tau_{\rm{tr}}},
\ee
and dimensionless energy $x=\epsilon/T$,  one can rewrite Eq.~\eqref{tau-tr-ph}
as follows:
\be
\gamma_{\rm{ph}} \sim  G  |x|^{2\eta -1} \times \left\{ \begin{array}{ll}
x^2/G_e,\ \ & \text{for}~~|x| \ll \sqrt{G_e}   \vspace{2mm} \\  1,\ \
& \text{for}~~|x| \gg \sqrt{G_e} .  \end{array} \right.
\label{gamma_ph}
\ee
Since $\gamma_{\rm{ph}} \propto |x|^{2\eta+1}$ for $|x| \to 0,$ the
conductivity  turns to infinity due to divergent contribution of small $x
:$ $\sigma_{\rm{ph}} \sim \int dx/|x|^{2\eta}$ [see Eq.~\eqref{sigma}].
Hence, surprisingly, when the screening is taken into account, the
electron-phonon scattering by itself is not sufficient to yield a finite
resistivity. Physically, this happens due to enhancement of screening with
approaching to the Dirac point, which results in shunting of the dc current by
low-energy electrons.  The divergency is cured by accounting of
electron-electron collisions whose rate scales at low energies as
\cite{Schuett-ee}
$\gamma_{\rm{ee}} \sim \sqrt{|x|}$. At low energies, such collisions
win competition with  electron-phonon scattering even for the case $G \gg  G_e$.
Consequently, a new low-energy scale, \be \epsilon \sim  T \left(\frac{G_e}{G}\right)^{2/(4\eta+1)} \ll T, \label{scale}\ee
appears in the problem, defined by the condition
$\gamma_{\rm{ee}} \sim  \gamma_{\rm{ph}}$.

\begin{figure}[ht!]
     \leavevmode \epsfxsize=8.0cm
 \centering{\epsfbox{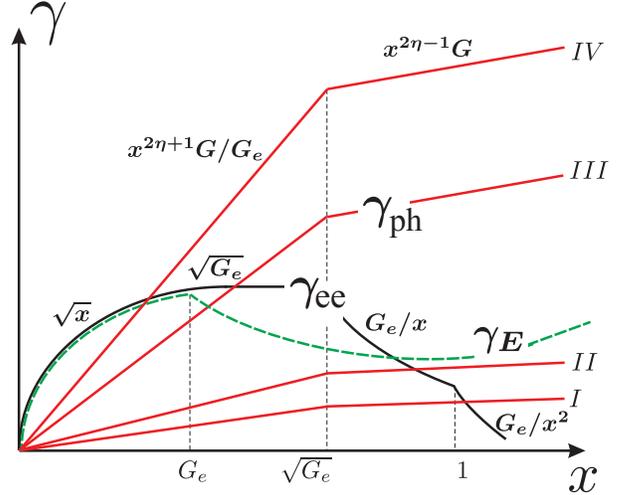}}
 \caption{Schematic plot of the dimensionless electron-phonon scattering rate
$\gamma_{\rm{ph}}(x)$ at $G_e\ll 1$  for
 four different values of $G$
  ($G$ increases from  \uppercase\expandafter{\romannumeral 1} to
\uppercase\expandafter{\romannumeral 4}). Also shown are the transport
scattering rate $\gamma_{\rm{ee}}(x)$ (thick solid) and the energy relaxation rate $\gamma_E(x)$ (dashed)
induced by electron-electron scattering.
}
\label{Fig1}
\end{figure}

Next, we analyze different transport regimes,
starting from the case $G_e\ll 1$. To this end, in
Fig.~\ref{Fig1}, we compare $\gamma_{\rm{ph}}(x)$  plotted schematically for $G_e\ll
1$ and four different values of coupling constant $G$  with dimensionless
electron-electron  transport and energy relaxation rates
[$\gamma_{\rm{ee}}(x)$ and $\gamma_E(x)$, respectively] calculated in
Ref.~\onlinecite{Schuett-ee}.  The averaging procedure appropriate for
evaluation of the conductivity depends on the relation between $\gamma_E$
and $\gamma_{\rm{ph}}+\gamma_{\rm{ee}}$. Specifically, for $\gamma_E \gg
\gamma_{\rm{ph}}+\gamma_{\rm{ee}}$ one should first average the total rate,
$\gamma_{\rm{ph}}+\gamma_{\rm{ee}}$,  over energy and use the thus obtained averaged rate
for calculation of conductivity [see Eqs.~ \eqref{tau-ave},
\eqref{Drude-competition0}, \eqref{tau-ave1},  and \eqref{Drude-competition}].  On the contrary, for
$\gamma_E \ll \gamma_{ph}+\gamma_{ee},$ the conductivity is controlled by
energy-averaged [see Eq.~\eqref{Drude}]   effective transport time,
$(\gamma_{\rm{ph}}+\gamma_{\rm{ee}})^{-1}.$ (In fact, the averaging procedure is only
important for the numerical coefficient.)

Let us consider   regimes \uppercase\expandafter {\romannumeral 1}--\uppercase\expandafter {\romannumeral 4} (see Fig.~1) realized with increasing
electron-phonon effective coupling $G$:
                \begin{itemize}
                \item  \uppercase\expandafter {\romannumeral 1}.   $G \ll G_e.$
Electron-phonon coupling is weak and $\gamma_{\rm{ph}} \ll \gamma_{\rm{ee}}$   within
relevant energy interval ($\epsilon <T,~ x<1 $), so that the phonon contribution
to the transport rate is negligibly small and $\Sigma=1/G_e.$ Relevant energies
are of the order of temperature, $x \sim 1.$
\item \uppercase\expandafter {\romannumeral 2}. $ G_e \ll G \ll G_e^{1-\eta}.$
Electron-phonon contribution dominates,  the screening of the phonons yields
negligible effect, implying that $\Sigma= 1/G$ and $x \sim 1$.
                \item \uppercase\expandafter {\romannumeral 3}. $G_e^{1-\eta} \ll G \ll G_e^{1/2-2\eta}.$ The same
                as for regime \uppercase\expandafter {\romannumeral 2,} $\Sigma=
1/G$,  $x \sim 1$.
                \item \uppercase\expandafter {\romannumeral 4}.
$G_e^{1/2-2\eta} \ll G. $ The conductivity is determined by a competition
between electron-electron collisions and screened electron-phonon interaction.
The dominant contributions comes from low energies,  $\ x\sim
(G_e/G)^{2/1+4\eta} \ll 1$, where $\gamma_{\rm{ph}} \approx \gamma_{\rm{ee}}$.  The
dimensionless conductivity $\Sigma$ scales with the coupling constants as
$\Sigma \sim (G_e/G)^{3/{1+4\eta}}.$
              \end{itemize}

Consider now the opposite case $G_e \gg 1.$  Calculations
analogous to the ones carried out in Ref.~\onlinecite{Schuett-ee} show that
in this case, $\gamma_{\rm{ee}} \sim \sqrt{x}$ for all relevant energies ($x<1$). In
the absence of phonons ($G=0$), the conductivity limited by electron-electron
collisions is given  by\cite{aleiner} $\sigma \sim e^2 N^2/ \hbar$,  and, consequently,
$\Sigma\simeq 1$. For $G \neq 0$ one should also take into account phonons
which are strongly screened  in
this case for all relevant energies, so that $\gamma_{\rm{ph}} \sim (G/G_e)
x^{2\eta+1}$ for $0<x<1$.  The phonon scattering becomes important when $G$
becomes larger than $G_e$. The main contribution to the resistivity comes
then from the region $ x\sim (G_e/G)^{2/1+4\eta} \ll 1$, where $\gamma_{\rm{ee}} \sim
\gamma_{\rm{ph}}$, yielding  $\Sigma \sim(G_e/G)^{3/{1+4\eta}}$.

\subsection{Results}
\label{s3.3}

\begin{figure}[ht!]
 \leavevmode \epsfxsize=8.0cm
 \centering{\epsfbox{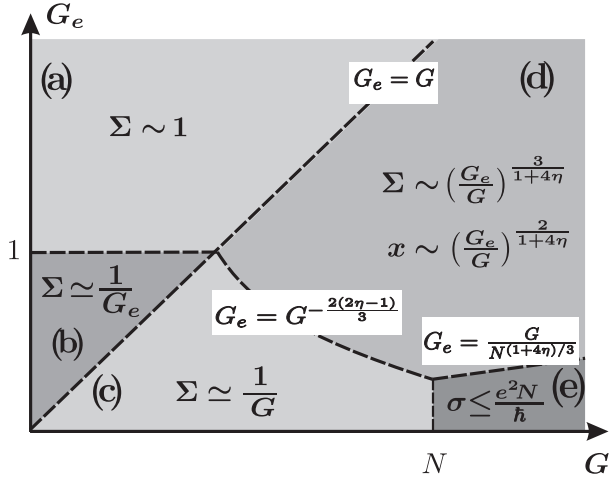}}
 \caption{Conductivity of clean suspended graphene at the Dirac point. Transport
regimes characterized by different behavior of dimensionless
conductivity $\Sigma$ are shown in the parameter plane of effective
(temperature-dependent) dimensionless couplings
$G$ and $G_e$. The analytical expressions for dashed lines separating    different regimes are given in the white boxes.
}
\label{Fig2}
\end{figure}

The above results are summarized in Fig.~\ref{Fig2}, which illustrates
different scattering regimes in the plane of parameters $G$ and $G_e$.
In the regions (a) and (b) the phonon scattering is weak and the conductivity is
limited by electron-electron collisions only. Contrary to this, in region (c)
the electron-electron interaction is weak, the phonons dominate transport
properties and their screening can be neglected. In the region (d), the
conductivity is determined by competition between electron-electron collisions
and scattering by screened phonons. As a result of this competition, a new
energy scale appears in the problem, where contributions of both types of
scattering are of the same order. Finally, on  the
boundary of the region (e), the conductivity  achieves its ``quantum limit''
of the order of $e^2N/\hbar.$  We expect that the conductivity saturates at
this value in the whole region (e).

\begin{figure}[ht!]
 \leavevmode \epsfxsize=8.0cm
 \centering{\epsfbox{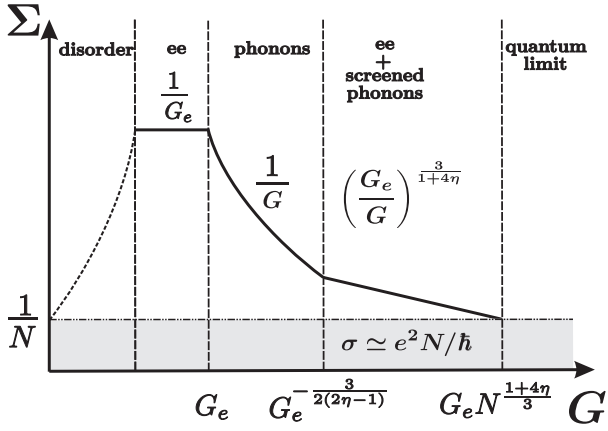}}
 \caption{ Dimensionless conductivity $\Sigma(G,G_e)$ at the Dirac point as a
function of
effective phonon coupling $G$ for fixed small electron-electron coupling $G_e$.
According to the definition of $G$, Eq.~(\ref{couplings}), the shown dependence can be
also understood as temperature dependence of the resistivity.
In the low-$G$ (low-temperature) region, the conductivity is governed by
impurity scattering (shown by dashed line).
}
\label{Fig3}
\end{figure}

It is worth reminding the reader  that all the temperature dependence has been
absorbed into dimensionless constants $G$ and $G_e$.
Since $G_e$ depends on $T$ in a very slow (logarithmic) manner, the dependence
of $\Sigma$ (and, consequently, $\sigma$) on $T$ is mostly determined by power-law
temperature dependence of $G$.
The  dependence of $\Sigma(G,G_e)$ on $G$ for fixed $G_e$ is illustrated  in
Fig.~\ref{Fig3} for the case of relatively small $G_e$ such that
$(1/N)^{2(2\eta-1)/3}<G_e<1$.  These inequalities correspond to a horizontal
line in $(G,G_e)$-plane (see Fig.~\ref{Fig2}) lying above the upper left
corner of the region $(e)$ but below $G_e=1$.   We see that at small $G$
(low temperatures) the electron-electron collisions dominate. At intermediate
temperatures $\Sigma$ is limited by phonons ($\Sigma \simeq 1/G$) and at high
temperatures the conductivity  is determined by  the narrow region of energies
where electron-electron collisions and scattering  on the  screened phonons have
approximately equal rates. In this region, $\Sigma\simeq
\left({G_e}/{G}\right)^{{3}/{1+4\eta}}$.  The perturbative  calculations
presented above become invalid at very high temperature, when $\Sigma$  drops
down to $\sim 1/N$ and, consequently, $\sigma$ becomes of the order of the
quantum limit $e^2N/\hbar$.

In order to make the picture complete, we also
showed in Fig.~\ref{Fig3} the contribution of a static disorder (if exists)
assuming that it is  due to randomly distributed charged impurities (another possible type of
disorder in suspended graphene---adatoms---would lead to similar results\cite{impurity}).
Such disorder dominates at low temperatures when $G$ is small.  Indeed,  transport
scattering rate due to charged impurities is inversely proportional to the
energy $1/\tau_{\rm{im}}(\epsilon)  \propto n_i/\epsilon$ (here $n_i$ is the impurity concentration) and at low $T$ exceeds electron-phonon
and electron-electron scattering rates at relevant energies $\epsilon \sim T.$
 The impurity-limited conductivity  is estimated as
$\sigma_{\rm{im}} \sim  (e^2 N/\hbar)   T \tau_{\rm{im}}(T) \propto T^2/n_i$.  This
equation is valid provided that $T \tau_{\rm{im}}(T) \gg \hbar,$ which implies that
temperature  is not too small.  With further lowering temperature,
$\sigma_{\rm{im}}$ saturates at   the value $e^2 N/\hbar$ (see
Refs.~\onlinecite{impurity,ostrovsky07,ostrovsky10} for the analysis
of the nature of the corresponding quantum critical point and discussion of related issues such as localization or antilocalization). As
illustrated in Fig.~\ref{Fig3}, impurity scattering becomes relevant when
$\sigma_{\rm{im}}$ becomes smaller than conductivity limited by phonons and
electron-electron collisions. The  decay of the conductivity both at
low and at high temperatures can be also
understood in the following way. Since the phonon potential is quasistatic,
it can be treated on equal footing with the impurity scattering.
Hence one could incorporate the impurity scattering into the  effective
coupling constant $G$, which would then become a non-monotonic function of
temperature $ G \propto T^{\eta} +1/T$ (we omit temperature-independent
coefficients). With decreasing temperature, ${G}$ defined in this way
would first fall, then reach the minimum, and then start to grow again, so
that both at very high and very low temperatures we would arrive at the region
(e) (``quantum limit'') in Fig.~\ref{Fig2}.

In the above analysis, we neglected contribution of other types of phonons. This
can be contrasted with the previous publications,
\cite{Mahan,Manez,CaKim,Oppen-short,Oppen-Scr}  where it was argued that the
contribution of phonon-induced random vector potential  might dominate over
deformation potential.    Let us estimate the contribution of the  gauge phonon
fields.   From Eqs.~\eqref{distortions} and \eqref{gauge}, one finds the
contribution of the flexural phonons to the gauge potential:
\be
A_x=\p_xh \p_y h,~~A_y=[(\p_xh)^2- (\p_yh)^2]/2.
\label{AA}
\ee
Similar to Eq.~\eqref{V}, gauge potential $V_\mathbf A$ is quadratic with respect to out-of-plane displacement of the graphene membrane.
Also, analogously to deformation potential,  the gauge field is quasistatic.
Hence the scattering rate may be calculated by using the golden rule for
scattering on the static potential $V_\mathbf A$.  Proceeding in this way, we
obtain
\be
\gamma_{\rm{ph}}^\mathbf A \propto G_\mathbf A x^{2\eta-1}
\label{gammaA}.
\ee
 Equation \eqref{gammaA} differs from the bottom line of Eq.~\eqref{gamma_ph}
only by replacement of $G$ with a much smaller constant:
 \be
 G_\mathbf A=G \left(\frac{g_2}{g_1}\right)^2\approx 2.5 \times 10^{-3} G\,,
 \label{G_A}
 \ee
so that the scattering off the gauge field is much less efficient than that off
the deformation potential. In fact, one should be slightly more careful at
this point, since, in contrast to the deformation potential, the gauge field is
not screened and Eq.~\eqref{gammaA} remains valid also at  $x <\sqrt{G_e}$,
where deformation potential scales as $1/G_e$ because of screening. Hence, with
increasing $G_e,$ gauge field may come into competition with deformation field.
Specifically, for
\be
G_e  \gg \left(\frac{g_1}{g_2}\right)^2\approx 4 \times 10^2,
\label{Ge_big}
\ee
one could neglect the latter contribution, and the conductivity would be limited
by a combined effect of the gauge field and electron-electron interaction.
Simple estimates show, however, that the inequality Eq.~\eqref{Ge_big} is hard
to satisfy in realistic system, especially when the logarithmic
renormalization of $g_e$ is taken into account.   Furthermore, there is a second
condition:
\be
G  \gg \left(\frac{g_1}{g_2}\right)^2,
\label{G_big}
\ee
which   ensures that the rate of electron-electron collisions is smaller than
the gauge-field scattering rate.  Only provided that both Eqs.~\eqref{Ge_big}
and \eqref{G_big} are satisfied (i.e. both $G$ and $G_e$ are very large), the
resistivity is controlled by the scattering off the gauge field,
$\Sigma=1/G_\mathbf A$. This situation appears highly unrealistic.
We thus conclude that the gauge phonon field does not essentially affect
transport properties of graphene
in the Dirac point up to very high and unrealistic values of $G$ and $G_e$.

One can also check that in-plane phonons do not
give essential contribution for realistic values of parameters.
This is consistent with the previous study \cite{Oppen-long}
that came to the same conclusion for the case of large chemical potential $\mu$
in the absence of externally applied strain.

\subsection{Phonon-induced velocity renormalization}
\label{s3.4}

In the previous sections we discussed  the electron scattering rate caused by
flexural phonons. Technically, this implied calculation of imaginary part of the
electron self-energy in the quasistatic phonon potential.   One may also
calculate the real part of the self-energy, thus extracting information about
a phonon-induced spectrum modification.  This will allow us to verify the
assumption that the electron spectrum is not changed essentially, which was
implicit in our perturbative analysis.
\begin{figure}[ht!]
 \leavevmode \epsfxsize=8.0cm
 \centering{\epsfbox{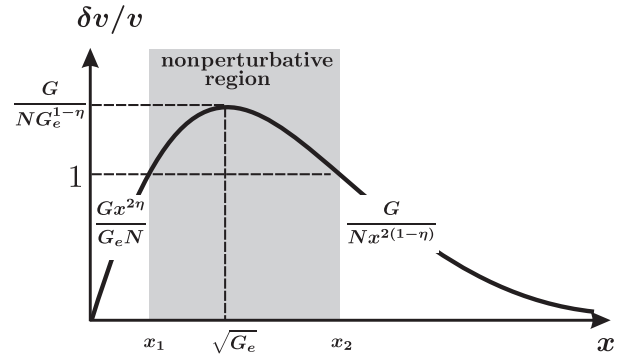}}
 \caption{ Velocity renormalization for $G_e\ll 1$  and $G/ N G_e^{1-\eta}>1$}
\label{Fig4}
\end{figure}

Simple calculations yield the following estimate for the energy-dependent
velocity renormalization, $\delta v,$ caused by  flexural phonons
damped by the Thomas-Fermi screening:
\be
\frac{\delta v}{v} \sim
\left\{ \begin{array}{ll} \displaystyle{\frac{G}{N G_e}
x^{2\eta}},\qquad \text{for}~~ x <\sqrt{G_e}, \vspace{2mm}  \\
\displaystyle{\frac{G}{N }\frac{1} {x^{2(1-\eta)}}},\qquad \text{for}~~ x
>\sqrt{G_e}. \end{array}\right.
\label{deltaV}
\ee
Comparing Eq.~\eqref{deltaV} with Eq.~\eqref{gamma_ph}, we see that
\be
\frac{\delta v}{v} \sim \frac{\hbar}{ \epsilon \tau_{\rm{tr}}},
\label{1/etau}
\ee
so that the relative correction to the velocity  is on the order
of the scattering-induced spectrum smearing.
\begin{figure}[ht!]
 \leavevmode \epsfxsize=8.0cm
 \centering{\epsfbox{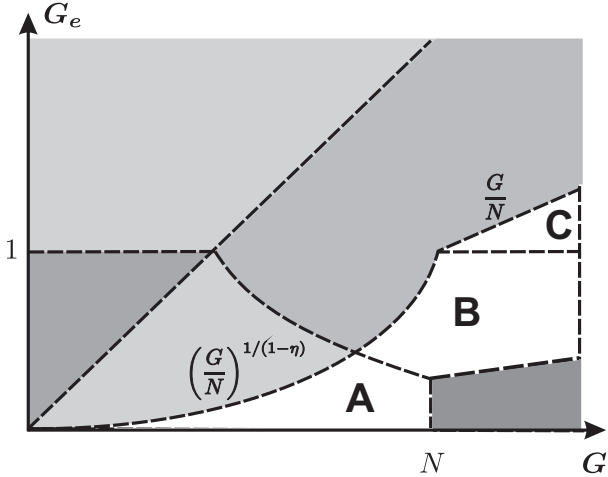}}
 \caption{  Regions A,B,C where   a ``non-perturbative'' interval of energies
exists }
\label{Fig5}
\end{figure}
In most of the cases, spectrum correction is small, $\delta v/v \ll 1,$ in the whole
relevant energy interval $x \leq 1$, and thus harmless. However, in certain
domains of parameters, the estimate~\eqref{deltaV} ceases to be small.
This may indicate that the calculation of the conductivity within the
lowest order of the perturbation theory (Born approximation) may become
insufficient.

Specifically, for $G_e>1$ we have    ${\delta v}/{v} \simeq ({G}/{N G_e})
x^{2\eta}$ at $x \leq 1$, so that the maximal value of   ${\delta v}/{v}$ is
achieved at  $x \sim 1$: ${\delta v_{\rm{max}}}/{v} \sim  {G}/{N G_e}$.  We thus
conclude that ``nonperturbative'' effects (i.e. those going beyond the Born
approximation) might show up for $G_e<G/N.$

 Consider now the opposite case $G_e<1,$ when the relative correction reaches the maximum
  \be
 \frac{\delta v_{\rm{max}}}{v} \sim \frac{G}{N G_e^{1-\eta}}
 \label{deltaVm}
 \ee
at $x\simeq \sqrt{G_e}.$  Hence, ``nonperturbative''  effects  come now into
play for $G_e<(G/N)^{1/(1-\eta)},$  as illustrated in Fig.~\ref{Fig4}.
From this estimates, we find the regions on the plane $(G,G_e)$  for which
$\delta v/v >1$ at some energy interval (but at the same time $\sigma
>e^2N/\hbar$). These regions are marked as  A,B, and C on Fig.~\ref{Fig5}.
In the regions A and B,  the velocity correction is not small in the
interval $x_1<x<x_2,$ where $x_1=(G_eN/G)^{1/2\eta}$ and
$x_2=(G/N)^{1/2(1-\eta)}$. One may expect that this ``nonperturbative energy
strip'' does not affect conductivity in the region A, because the width of this
strip $x_2-x_1$ is much smaller than the temperature window: $ x_2-x_1 \ll 1$.
Analogously, the perturbative analysis is expected to give the correct result in
region C, because $\delta v/v \ll 1$ in the range $x\sim
(G_e/G)^{2/(1+4\eta)} ,$ which is determined by the condition $\gamma_{\rm{ph}} \sim
\gamma_{\rm{ee}}$ and governs the resistivity within the perturbative calculation
(see previous section). On the other hand, a controlled calculation
of the conductivity  in the region B requires going beyond the Born
approximation.

The discussion of such ``nonperturbative'' phenomena is out of scope of the
current work and we restrict ourselves to a short comment of a somewhat
speculative character. One may expect that the system becomes strongly
inhomogeneous, i.e., it can be characterized by a local chemical potential
$\mu(\mathbf r)$ showing large fluctuations around zero (with large number of
electrons within each such ``puddle'').
Such fluctuations  can be described in the framework of random resistor network,
which involves percolation physics.\cite{cheianov}
Also,
within a spatial scale where $\mu(\mathbf r)$  is homogeneous, the spectrum  of
electrons and holes might be essentially different from the linear one.\cite{comment}
This and related  issues will be discussed elsewhere.

\section{Away from the Dirac point: \ \ finite $\mu$}
\label{s4}

In Sec.~\ref{s3}, we considered conductivity at the Dirac
point ($\mu=0$), which is in the main focus of this paper. In the present
section, we briefly discuss the behavior of conductivity away from the Dirac
point, $\mu\neq 0$. The phonon-limited resistivity in this regime has been
previously analyzed in Refs.~\onlinecite{Oppen-long} and \onlinecite{Ochoa} where the
renormalization of $\varkappa$ was neglected. While this is justified in the
presence of sufficiently strong externally induced tension, the renormalization
essentially affects the scattering rate when tension is absent (or weak), see
Sec.~\ref{s2}. Below we explore the effect of flexural phonons on resistivity of
graphene at nonzero $\mu$ in the absence of tension, with taking into account
the anharmonic renormalization.

First, we ignore the electron-electron interaction.
Substituting  Eq.~\eqref{tau-tr-appendix} into Eq.~\eqref{sigma}, we get
\BEA
\sigma_{\rm{ph}}&= & \frac{e^2}{\hbar}\frac{N}{g^2}\label{sigma1} \\ \nonumber &\times& \left\{ \begin{array}{ll}
\displaystyle{\frac{ C_1}{16 Z^2
C}\left(\frac{\Delta_c}{T}\right)^\eta },\ \  & \text{for} ~|\mu| \ll T,
\vspace{2mm} \\ \displaystyle{\frac{1}{8C Z^2}
\left(\hspace{-1mm}\frac{\Delta_c}{T}\hspace{-1mm}\right)^\eta
\hspace{-2mm}\left(\frac{\mu}{T}\right)^{2-2\eta} }\hspace{-2mm},\ \ &
\text{for} ~T \ll \hspace{-1mm}|\mu| \hspace{-1mm}\ll \sqrt{T\Delta_c},
\vspace{2mm}  \\ \displaystyle{\frac{1}{8}
\left(\frac{\mu}{{T}}\right)^2 \frac{1}{\ln(\mu/\sqrt{T\Delta_c})} },\
\ & \text{for} ~  \sqrt{T\Delta_c} \ll |\mu| \,.
 \end{array} \right.
\EEA
We see
that the conductivity increases with $\mu $  as $\mu^{2-2\eta}\simeq \mu^{0.56}$
for
$T \ll \mu \ll \sqrt{T\Delta_c}$ and as $\mu^2/\ln\mu$ for
$   \sqrt{T\Delta_c} \ll \mu.$

 Now we include the Coulomb interaction into consideration. Just as in
the case $\mu=0$ (Sec.~\ref{s3}),  its role is twofold: first, it screens
the deformation potential and, second, it opens an additional
(electron-electron) channel of scattering.

For $\mu \ll T$ [first line  in  Eqs.~\eqref{sigma1}], the  effect of the
screening was discussed in the previous sections
[see Eqs.~\eqref{TF2}--\eqref{tau-tr-Q2} and \eqref{tau-tr-ph}].
For $\mu \gg T,$ the Thomas-Fermi screening leads to the renormalization of $g$
(see Ref.~\onlinecite{Oppen-long})  described by Eq.~\eqref{TF4} with
\be
\Pi(Q)=\frac{\mu}{2\pi \hbar^2 v^2}.
\label{TF3}
\ee
Thus, we have to replace
\be
g \to \frac{g}{ 1+g_e N k_F/ Q },
\label{TF1}
\ee
where $k_F=\mu/\hbar v.$  The transport scattering rate is calculated in  Appendix \ref{a1-3}. For $G_e \ll 1 $ screening can be neglected so that $1/\tau_{\rm{tr}}$ and $\sigma_{\rm{ph}}$ are given by   Eqs.~\eqref{tau-tr-appendix} and \eqref{sigma1}, respectively.  For $G_e \gg 1 ,$ scattering rate reads
\BEA
\frac{1}{\tau_{\rm{tr}}(\epsilon)}&\simeq&
\frac{2 g^2T^2|\epsilon|}{\pi \hbar G_e\mu^2}\label{tau-tr-appendix-scr}
 \\  \nonumber &\times& \left\{ \begin{array}{ll} \displaystyle{
C_3\ln\left(\frac{|\epsilon|}{\sqrt{T\Delta_c}} \right) },\ \ & \text{for}
~~|\epsilon| \gg \sqrt{T\Delta_c}, \vspace{2mm}  \\ \displaystyle{ \tilde{C}
Z^2~\left( \frac{|\epsilon|}{\sqrt{T\Delta_c}}\right)^{2\eta}},\ \ & \text{for}
~~|\epsilon| \ll \sqrt{T\Delta_c}\,,
\end{array} \right.
\EEA
where $\tilde{C}\approx 4.$
Then, for
$G_e \gg 1,$ the conductivity limited by screened flexural phonons takes the
form
\BEA
\sigma_{\rm{ph}}&\simeq&\frac{e^2}{\hbar}\frac{N G_e}{g^2} \label{sigma-mu}
\\ \nonumber
&\times& \left\{ \begin{array}{ll}
\hspace{-2mm}\displaystyle{\frac{1}{8 \tilde{C} Z^2}
\left(\hspace{-1mm}\frac{\Delta_c}{T}\hspace{-1mm}\right)^\eta
\hspace{-1mm}\left(\hspace{-1mm}\frac{\mu}{T}\hspace{-1mm}\right)^{2-2\eta} }
\hspace{-2mm},~\text{for} ~T \ll \hspace{-1mm}|\mu|
\hspace{-1mm}\ll \hspace{-1mm}\sqrt{T\Delta_c}, \vspace{2mm}
\\ \hspace{-2mm} \displaystyle{\frac{1}{8}
\left(\hspace{-1mm}\frac{\mu}{{T}}
\hspace{-1mm}\right)^2 \frac{1}{C_3\ln(\mu/\sqrt{T\Delta_c})} },~\text{for}
~  \sqrt{T\Delta_c} \ll |\mu|. \end{array} \right.
\EEA

\begin{figure}[ht!]
 \leavevmode \epsfxsize=8.0cm
 \centering{\epsfbox{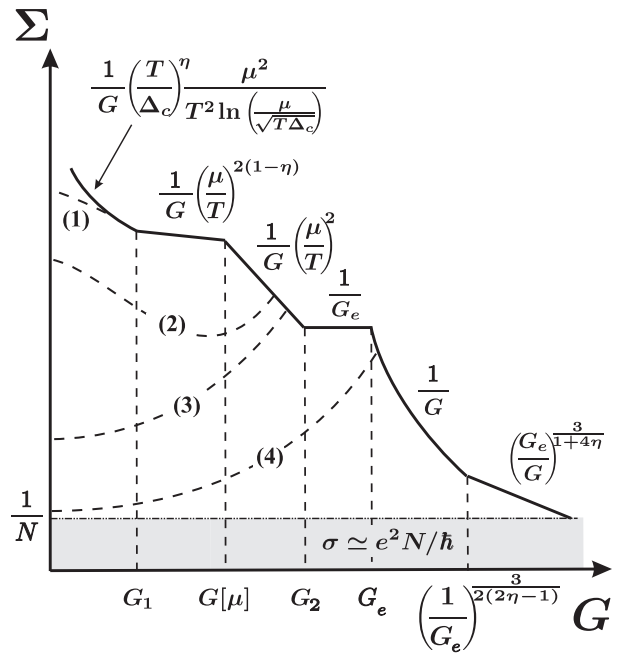}}
 \caption{ Temperature dependence of dimensionless conductivity $\Sigma$ for
$G_e \ll 1$ and relatively small $\mu$ such that $G[\mu] \ll G_e$. Dashed lines
show    conductivity limited by  charged impurities for different impurity
concentration increasing from the curve (1) to the curve (4).    }
\label{Fig6}
\end{figure}

Finally, we discuss the role of electron-electron collisions at $\mu \neq 0$.
General equations describing the competition between  electron-electron
collisions and electron-phonon scattering   for arbitrary $\mu$ are derived
in Appendix B. As follows from these results, for $\mu \gg T$
electron-electron collisions do not contribute to effective scattering rate
 and the conductivity is given by Eq.~\eqref{sigma-mu}. This implies that the
electron-phonon scattering becomes  even more important when the chemical
potential is tuned away from the Dirac point. This conclusion is supported by
consideration of the conductivity in the region $0<\mu \ll T$ for the case, when
electron-electron collisions dominate at $\mu=0.$  As shown in Appendix B, the
conductivity is given in this case by
\be
\sigma \sim \frac{e^2 N T}{\hbar}\langle 1 \rangle \left[\frac{1}{\langle 1/\tau_{\rm{ee}}\rangle} + \frac{\mu^2}{T^2} \frac{1}{\langle 1/\tau_{\rm{ph}} \rangle }\right].
\label{small-mu}
\ee
(here we omit numerical coefficients of order unity).
We see that the electron-phonon scattering, being weak at $\mu=0$, becomes
nevertheless  dominant at a quite small chemical potential  $\mu \sim T
\sqrt{\langle 1/\tau_{\rm{ph}} \rangle /\langle 1/\tau_{\rm{ee}}\rangle} \ll T.$
The behavior of the  dimensionless conductivity $\Sigma$ with
increasing temperature for $\mu \neq 0$ is shown schematically in
Fig.~\ref{Fig6}. It is  assumed  in this figure that  $\mu$ is relatively
small, so that  the temperature-dependent coupling constant $G=G[T]$ is weak for
$T\sim \mu$ as compared to the coupling constant $G_e[T]$ (which depends only
logarithmically weakly on temperature),
\be
G[\mu]\ll G_e[\mu]\,.
\label{ineq2}
\ee

For
  $G[T] \ll G_2 $ (here $G_2=G[T_2]$ and
$T_2\gg\mu$ should be  found from the equation $G[T_2]=G_e\mu^2/T_2^2$)
  the conductivity is limited by screened phonons.  For very
small $G$ such that $G \ll G_1=G[\mu^2/\Delta_c],$  the phonons can be treated
in the harmonic approximation.  For $G \gg G_2$ the temperature dependence is
the same as in Fig.~3.   At lowest temperatures the conductivity is limited by
the scattering  off static disorder (charged impurities) as illustrated in
Fig.~\ref{Fig6} by dashed lines corresponding to different impurity
concentrations. Indeed, the phonon transport rate,
Eq.~\eqref{tau-tr-appendix-scr},  taken at $\epsilon =\mu,$ has a maximum as a function of $\mu$
at $\mu \approx \sqrt{T \Delta_c}$ and the maximal value decreases  with decreasing the temperature as
  $T^{3/2}.$ Therefore, at sufficiently  low temperatures
scattering by charged impurities with the temperature-independent  rate, $1/\tau_{\rm{im}} \propto
n_{i}/|\epsilon| \approx n_{i}/\mu,$ dominates within the relevant energy interval:
 \be      \frac{1}{\tau_{\rm{im}}(\epsilon)} \gg \frac{1}{\tau_{\rm{tr}}(\epsilon,T)}, ~~\text{for}~|\epsilon-\mu| \lesssim T.
 \label{ineq3}
 \ee
   Here $\tau_{\rm{tr}} (\epsilon, T )$ is the phonon-induced transport time given by
Eq.~\eqref{tau-tr-appendix} or Eq.~\eqref{tau-tr-appendix-scr}; it is
convenient for our purposes  to write it as a function of two variables
$\epsilon$ and $T.$
The impurity scattering time $\tau_{\rm{im}}$   is in fact also  a function of two independent variables $\epsilon$ and $T$ because of screening of the impurity potential. This effect does not change qualitatively our results, so that  we do not  discuss it  here.

Let us find the temperature behavior of the
impurity-dominated conductivity   assuming that Eq.~\eqref{ineq3} is satisfied. As
discussed above,  electron-electron collisions  become irrelevant  while going away from the Dirac point,  so that we neglect them and write the conductivity as
      \BEA
&&\sigma= \frac{e^2 N}{16\pi\hbar^2} \times \label{sigma-im}\\ \nonumber &&\int_{-\infty}^\infty d \epsilon \frac{|\epsilon| }{T \cosh^2[(\epsilon-\mu)/2T]} \frac{\tau_{\rm{tr}}(\epsilon,T)\tau_{\rm{im}}(\epsilon)}{\tau_{\rm{im}}(\epsilon) +\tau_{\rm{tr}}(\epsilon,T)}.
\EEA
 Using Eq.~\eqref{ineq3} we expand Eq.~\eqref{sigma-im} over power of $\tau_{\rm{im}}/\tau_{\rm{tr}}$ and keep terms of the zero and first order:
 \be\label{sigma-im1}
\sigma=\frac{e^2 N}{16\pi\hbar^2}  \int_{-\infty}^\infty d \epsilon \frac{|\epsilon|\left[\tau_{\rm{im}}(\epsilon) -{\tau_{\rm{im}}^2(\epsilon)}/{\tau_{\rm{tr}}(\epsilon,T)}\right] }{T \cosh^2[(\epsilon-\mu)/2T]} .
\ee
 Since $T\ll\mu,$ the integrand  is peaked near the region $\epsilon \approx \mu.$   While calculating  contribution of the first  term in the square brackets we write $ |\epsilon| \tau_{\rm{im}}(\epsilon) \approx \mu \tau_{\rm{im}}(\mu) + [\mu \tau_{\rm{im}}(\mu)]' (\epsilon-\mu)+ [\mu \tau_{\rm{im}}(\mu)]'' (\epsilon-\mu)^2/2.$ The second term in the square brackets of Eq.~\eqref{sigma-im1} is small and while integrating it we neglect temperature broadening of the Fermi-function. Doing so, we find for the temperature-dependent part of the conductivity,
 \BEA
  \delta \sigma &=& \sigma(T)-\sigma(0) \label{dsigma} \\ \nonumber  &=& \frac{e^2 N}{4\pi\hbar^2}\left\{ \frac{\pi^2T^2}{6}[\mu\tau_{\rm{im}}(\mu)]'' -\frac{\mu \tau_{\rm{im}}^2(\mu)}{\tau_{\rm{tr}}(\mu,T)}\right\}.
 \EEA
 This equation is valid for an arbitrary type of impurity scattering. For
charged impurities we find that
    $\delta\sigma $ is given as a sum of two terms of different signs:
 \be \delta \sigma(T) \propto  \frac{T^2}{n_{i}}-\frac{\mu^3}{n_{i}^2}\frac{1}{\tau_{\rm{tr}}(\mu,T)}\label{sigma-T}\ee
  (here we omit coefficients which do not depend on $\mu,T$ and $n_i$). The
analysis of Eq.~\eqref{sigma-T} shows that   ``metallic'' behavior of
conductivity (i.e. its increase with lowering $T$) at  low $n_{i}$ changes to
an ``insulating'' one with increasing $n_{i}$ as illustrated in Fig.~\ref{Fig6}
by dashed lines. Alternatively, this crossover in the behavior of
conductivity at relatively lower temperatures can be observed if one changes
$\mu$ at fixed disorder. One can also see from Fig.~\ref{Fig6} that at
intermediate values of $n_{i}$ (or of $\mu$) the dependence of conductivity on $T$
may have two maxima [see curve (2) in Fig.~\ref{Fig6}].

\section{Comparison to experiment}
\label{s5}

We now
compare our results  with available experimental
data \cite{Bolotin} (see also Ref.~\onlinecite{Ochoa}).
To this end,       we plot conductivity as a  function of  temperature  and  chemical potential (or, equivalently,  electron concentration) for the same values of parameters as
in Ref.~\onlinecite{Bolotin}.  We assume that electron-electron coupling is renormalized  from the value $g_e^0$  at
large energies ($\sim \Delta$) to $g_e \simeq 1$ in the room temperature interval, so that  $G_e=C_3g_e^2N^2\simeq 16 \gg 1 $.   The
conductivity is obtained   by interpolating of equation  for total scattering rate (which is the sum of  the electron-electron,
electron-phonon, and impurity scattering rates)  between asymptotical expressions presented in the previous sections and by substituting
thus found rate  into the  Drude formula Eq.~\eqref{Drude}.   The main purpose  of such an interpolation is to   estimate characteristic values of conductivity
and analyze qualitatively the dependence of $\sigma$ on $\mu,T$ and $n_i.$
Due to
evident reasons, we do not pretend   to get quantitative agreement with experiment.  First of all, interpolation procedure
yields numerical value of conductivity up to the coefficient on the order of unity at the boundaries separating    regions  of parameters corresponding to  different
transport regimes. Second,  even asymptotical expressions
contain
some  unknown numerical coefficients. In particular, the coefficient  $Z \sim 1$ entering the phonon correlation function [see Eq.~\eqref{G-eta}] is not known as it was discussed above. Below we use   $Z =2 ,$ which allows us to get reasonable agreement with experiment. We also do not know  the numerical coefficient in the  equation $\gamma_{\rm{ee}}\sim \sqrt{x}$ for dimensionless electron-electron scattering rate in the limit $G_e \gg 1.$ In the estimates we choose this coefficient to be $1/4$ which yields a best fit to experimental data.  Another issue, which was not resolved rigourously, is the screening of the impurity potential.  In the estimates we simply assume $1/\tau_{\rm{im}} \sim e^4n_i/|\epsilon|\kappa$  and choose  the coefficient in this equation to be unity. This yields a good approximation for impurity scattering rate (up to a numerical coefficient) at least for not too low temperatures, when conductivity is much larger than $e^2/h.$

\begin{figure}[ht!]
     \leavevmode \epsfxsize=8.0cm
 \centering{\epsfbox{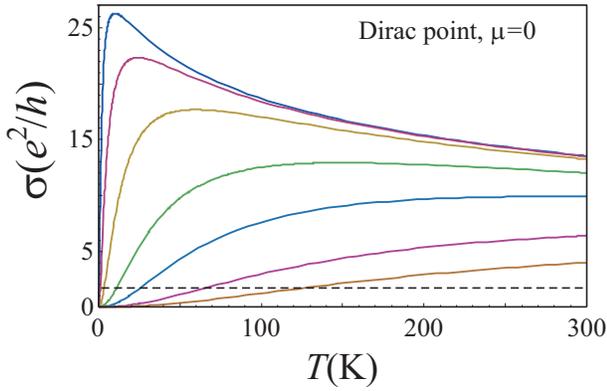}}
 \caption{ Conductivity at the Dirac point ($\mu=0$) for different values of the impurity concentration
 ($n_i/10^{10}~ \text{cm}^{-2}= 10^{-4},~ 10^{-3},~10^{-2},~10^{-1},~ 0.5,~ 3 ,~10$) increasing from the top to the bottom.  Dashed line
 corresponds to    SCBA limit  $\sigma= 4e^2/\pi h .$   Calculations are  controlled only well above this line.
       }
\label{Fig7}
\end{figure}

At the Dirac point
the data shown in Fig.4 of Ref.~\onlinecite{Bolotin} show an increase of
conductivity by factor of $2 \div 3$ when temperature increases from 5  to 200
K, with a clear saturation around 200 K. Let us compare this picture  with theoretical estimates.
The  calculated conductivity at the Dirac point is plotted in Fig.~\ref{Fig7} for different values
of impurity concentration $n_i$.
A  curve corresponding to $n_i=0.5 \times 10^{10} \text{cm}^{-2}$
(third curve counted from the bottom)  reasonably agrees  with experiment.
A similar behavior with a somewhat larger ratio $\sigma(200\:{\rm
K})/\sigma(5\:{\rm K})$ is seen in Fig.~3 of Ref.~\onlinecite{Ochoa}. We expect
that  with further increase of temperature the conductivity
will drop due to electron-phonon scattering, as found in Sec.~\ref{s3} of
the present work and is shown schematically  in
our Fig.~\ref{Fig3}.  As seen from Fig.~\ref{Fig7}, the drop  of the conductivity with $T$ can be observed at smaller $T$ provided that   one uses cleaner samples.

Let us note that the conductivity curves in   Fig.~\ref{Fig7}  go to zero in the limit
$T\to 0,$  in view of the vanishing density of states at the
Dirac point. If we would include disorder-induced level broadening
self-consistently [in the framework of the self-consistent Born
approximation (SCBA)], we would get instead a limiting conductivity value $
4e^2/\pi h$ (marked by a horizontal dashed line in the plots). The actual behavior
of conductivity in this regime is controlled by quantum interference effects
that lead to localization, antilocalization, or quantum criticality, depending
on the character of disorder.\cite{ostrovsky07}  In the present
paper we do not discuss these phenomena, as our focus is on regimes where the
dimensionless conductivity is sufficiently large and quantum interference
corrections do not change it significantly.

\begin{figure}[ht!]
     \leavevmode \epsfxsize=8.0cm
 \centering{\epsfbox{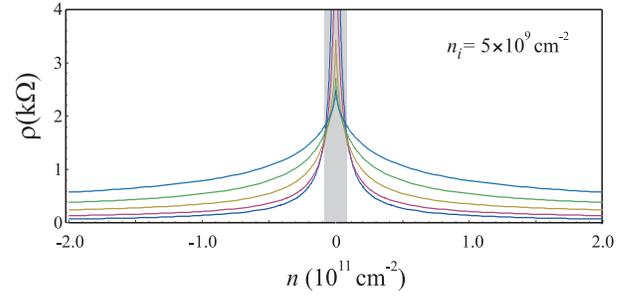}}
 \caption{ Resistivity as a function of electron concentration at  $n_i= 5 \times 10^9~ \text{cm}^{-2}$  and different temperatures ($T/1\text{K}=5,~40,~90,~150,~230$) increasing from the bottom to the top at large $n$.
 Within the   grey  area  temperature dependence is ``insulating'', while outside this region it is ``metallic''.    }
\label{Fig8}
\end{figure}

 When one moves away from the Dirac point, the experimentally observed
temperature dependence changes (see Fig.2 of Ref.~\onlinecite{Bolotin}), and at
sufficiently large $\mu$ the conductivity becomes monotonously decreasing
function of $T$. This evolution is in a very good  qualitative agreement with our results.
To see this, we fixed the impurity concentration at the level  $n_i=0.5 \times 10^{10} \text{cm}^{-2}$, and calculated
resistivity (just as in  Fig.2 of Ref.~\onlinecite{Bolotin}) as a function of the electron concentration $n=N\mu^2/4\pi\hbar^2v^2$  for different
values of temperature (the same as used in Ref.~\onlinecite{Bolotin}). The results are plotted in Fig.~\ref{Fig8} and look very similar to the ones presented in  Fig.2 of Ref.~\onlinecite{Bolotin}.
\begin{figure}[ht!]
     \leavevmode \epsfxsize=8.0cm
 \centering{\epsfbox{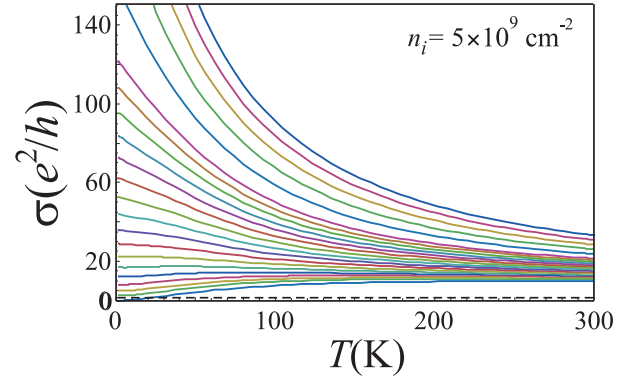}}
 \caption{ Conductivity at fixed impurity concentration ($n_i=5 \times 10^9~\text{cm}^{-2}$) for different values of chemical potential
 ($\mu/1\text{K}=0$, 50, 70, 90, 110, 130, 150, 170, 190, 210, 230, 250, 270, 290, 310, 330, 350, 400, 450, 500, 550, 600)
 increasing from the  bottom to the top.  Dashed line   corresponds to    SCBA limit  $\sigma= 4e^2/\pi h .$  }
\label{Fig9}
\end{figure}
\begin{figure}[ht!]
     \leavevmode \epsfxsize=8.0cm
 \centering{\epsfbox{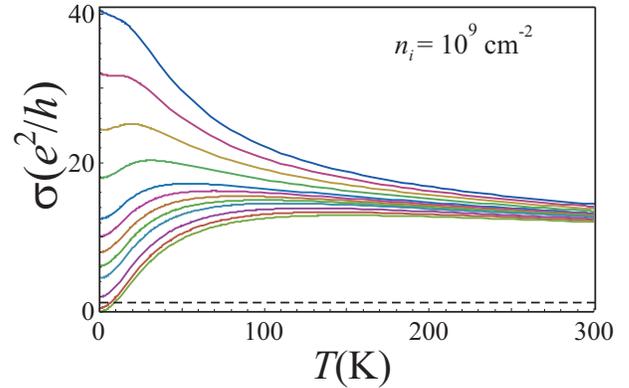}}
 \caption{ Conductivity at fixed impurity concentration ($n_i=10^9~\text{cm}^{-2}$) for different values of chemical potential
 ($\mu/1~ \text{K}=0,~ 10,~ 20,~30,~35,~40,~45,~50,~60,~70,~80,~90$)
 increasing from the  bottom to the top.  Dashed line   corresponds to    SCBA limit  $\sigma= 4e^2/\pi h .$ }
\label{Fig10}
\end{figure}
%
Within the   grey  area  temperature dependence is ``insulating'', while outside this region it is ``metallic''.    One of  the main features
 of this picture is the existence of the ``stationary'' point, where  $T$-dependence  changes.
 To illustrate the existence of this point in a more transparent  way we  plotted in Figs.~\ref{Fig9} and \ref{Fig10} the
 conductivity as a function of temperature for fixed impurity concentration and different values of the chemical potential.
The ``stationary'' point in Fig.~\ref{Fig8} corresponds to existence of more or less horizontal lines in
Figs.~\ref{Fig9} and \ref{Fig10} separating  regions with ``metallic''  and   ``insulating'' behavior.
As seen from Figs.~\ref{Fig9} and \ref{Fig10}, the transition between different types of T-dependence becomes
more pronounced with decreasing the impurity concentration.   It is worth noting that the transition may be also
obtained for fixed $\mu$ by changing the impurity concentration (for example, by annealing the sample) as illustrated in Fig.~\ref{Fig11}.
\begin{figure}[ht!]
     \leavevmode \epsfxsize=8.0cm
 \centering{\epsfbox{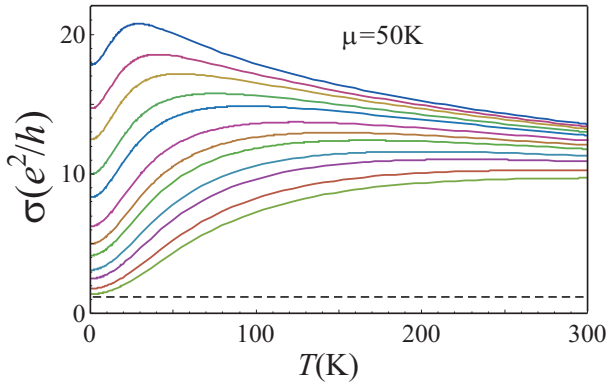}}
 \caption{  Conductivity at $\mu=50$~K for different values of the impurity concentration
 ($n_i/10^{10}~ \text{cm}^{-2}= 0.07,~ 0.085,~ 0.1,~ 0.125,~0.15,~0.2,~0.25,~0.3,~0.4,~0.5,~0.7,~0.9$) increasing from the top to the bottom.
 Dashed line   corresponds to    SCBA limit  $\sigma= 4e^2/\pi h .$}
\label{Fig11}
\end{figure}

Sufficiently far from the Dirac point,
such that $\mu \gg T$ for all relevant temperatures, only the first two
phonon-controlled regimes of Fig.~\ref{Fig6} survive, implying a crossover from
the conductivity scaling $\sigma \propto \mu^2/T^2 \ln(\mu^2/T\Delta_c)$ at
lower temperatures ($T \ll \mu^2/\Delta_c$) to $\sigma \propto
\mu^{2-2\eta}/T^{2-\eta}$ at higher temperatures $T\gg \mu^2/\Delta_c$.
However, the temperature, separating two regimes, $\mu^2/\Delta_c,$ turns out to be very small, on the order of $1$ K, even for largest chemical potentials, $\mu \simeq 600$ K, studied in
 Ref.~\onlinecite{Bolotin}. In other words, experimental situation  corresponds to anharmonic regime.
 While the data shown in Fig.~3 of Ref.~\onlinecite{Bolotin} do indicate a
power-law  increase of resistivity with temperature (at relatively large temperature), the
exponents do not quite agree: the theoretical dependence, $\rho \sim T^{2-\eta} \sim T^{1.3}$ [see upper line of Eq.~\eqref{sigma-mu}], turns out to be slightly stronger  than experimentally observed linear one,   $\rho \sim  T.$

\begin{figure}[ht!]
 \leavevmode \epsfxsize=8.0cm
 \centering{\epsfbox{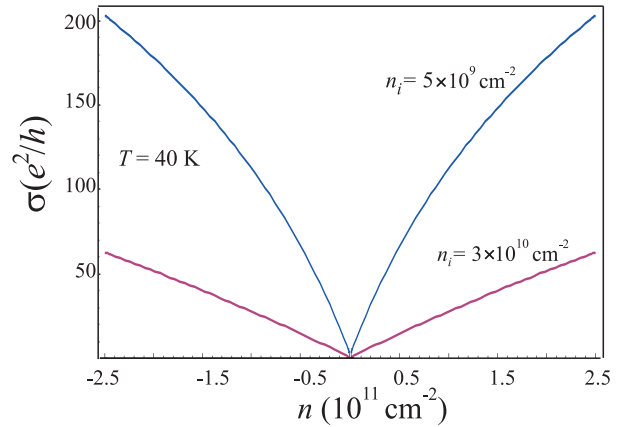}}
 \caption{ Conductivity as a function of electron concentration at $T=40$~K and two different values of impurity concentration:
 $n_i= 5 \times 10^9~ \text{cm}^{-2}$  and $n_i=3 \times 10^{10}~ \text{cm}^{-2}.$}
\label{Fig12}
\end{figure}

Consider now the dependence of the conductivity   on the electron density. As observed in  Ref.~\onlinecite{Bolotin}, this dependence is
qualitatively different in clean and dirty samples.  When impurity concentration is large, conductivity is a linear function of the concentration, $\sigma \sim n,$ while at the same sample  after annealing the dependence becomes  sublinear: $\sigma$
increases with  electron density $n
\propto \mu^2$ with an exponent considerably smaller than unity, which is at
least in qualitative agreement with the
$\sigma \propto
\mu^{2-2\eta} \propto n^{1-\eta}$ predicted above for transport away from the Dirac point [see upper line of Eq.~\eqref{sigma-mu}]. This is illustrated in Fig.~\ref{Fig12}, where $\sigma$  is plotted as function of $n$ for fixed temperature, $T=40$ K (the same as in Ref.~\onlinecite{Bolotin}), and two different values of the impurity concentration. This picture looks very similar to the Fig.~1  in Ref.~\onlinecite{Bolotin}. Physically, the linear dependence at large $n_i$ is caused by impurity scattering, while sublinear one at low $n_i$ is due to the phonon scattering.

%

Finally, we make a rough estimates of the
numerical values of the conductivity. For this purpose, we
consider temperature dependence of the conductivity for    $n_i=0.5 \times 10^{10} \text{cm}^{-2}$ and different values of $\mu$ (see Fig.~\ref{Fig9}).
For
$\mu=600\:{\rm K}$ (which corresponds to the highest density $n_e=2\times
10^{11}\:{\rm cm}^{-2}$ studied in Ref.~\onlinecite{Bolotin}) and $T=200\:{\rm
K}$
(which is approximately the highest temperature in Ref.~\onlinecite{Bolotin})
theoretical prediction is $\sigma
\simeq 50 e^2/h $ (upper curve in Fig.~\ref{Fig9}). In the Dirac point theory predicts $\sigma
\simeq 9 e^2/h $  (lowest  curve in Fig.~\ref{Fig9}). The experimental data of Ref.~\onlinecite{Bolotin} yield $\sigma\simeq 70
e^2/h $ for $\mu=600\:{\rm K}$  and $\sigma
\simeq 12 e^2/h $  for $\mu=0$, in reasonable agreement with our findings.

It was suggested in Refs.~\onlinecite{Oppen-long,Ochoa} that the experimental
data of Refs.~\onlinecite{Bolotin,Ochoa} might be to some extent affected by
strain that may result from fixing the sample at the contacts. It is worth
emphasizing that in the present paper we have studied the case of a strain-free
graphene and obtained reasonable qualitative agreement  with experiment.
Furthermore, the strength of the tension depends on the procedure of
preparation of a sample. In particular, the built-in tensions in freely
suspended graphene monolayers produced by means of chemical reduction of
graphene oxide were found~\cite{tension} to be considerably
weaker than in mechanically exfoliated graphene samples.

\section{Summary}
\label{s6}

To conclude, we have studied transport in suspended clean graphene in a broad range
of temperatures. We have explored the interplay of electron-phonon and
electron-electron interactions and have found that the scattering off flexural
phonons controls the resistivity at relatively high  $T.$ Taking
into account the anharmonic coupling of flexural and in-plane phonons was
crucial for correct evaluation of the graphene conductivity.

Our results for the conductivity can be expressed in terms of two
dimensionless coupling constants $G$ and $G_e$, characterizing the strength of
the electron-phonon and electron-electron scattering, respectively. Both
constants depend on temperature: $G$ shows a power-law scaling,   $ G \sim
T^\eta$,  with the exponent $\eta$ describing the scaling of the
bending rigidity of graphene membrane with the length scale due to
anharmonicity, while $G_e$ slowly (logarithmically) changes with $T$ due to
renormalization of the Fermi velocity caused by electron-electron interaction.

At the Dirac point, the dimensionless conductivity $\Sigma(G,G_e)$ of clean
suspended graphene depends on $T$ only through temperature  dependence of $G$
and $G_e.$ At low temperatures, $G<G_e$ and phonon scattering is negligible.
In the room temperature interval $G \gg G_e$  and transport is dominated by
phonon scattering. At sufficiently high temperatures one should take into
account screening of the deformation potential due to electron-electron
interaction.
The overall ``phase diagram'' of scattering regimes and conductivity scaling of
clean graphene at the Dirac point is shown in Fig.~\ref{Fig2}. A characteristic
temperature dependence of conductivity is shown in Fig.~\ref{Fig3}. There, we
also included the regime of low temperatures where the  resistivity is
controlled by disorder. [The lowest temperatures where quantum
criticality or localization (antilocalization) effects come into play are not considered
in this paper, see Refs.~\onlinecite{impurity,ostrovsky07,ostrovsky10}.] As seen
in this figure, the conductivity first increases with $T$ (disorder-dominated
regime), then shows a plateau (electron-electron scattering), and then drops
due to electron-phonon scattering. In the high-temperature part of the latter
regime, the Thomas-Fermi screening of phonons becomes important.  Remarkably, at
still higher temperatures the system enters the ultimate quantum regime
(conductivity of the order of $e^2/h$ analogous to the one at lowest
temperatures). In view of the quasistatic nature of flexural phonons, quantum
interference phenomena are expected to be relevant in this regime, despite
rather high temperatures.

Away from the Dirac point ($\mu \ne 0$),  the role of the electron-phonon interaction
increases as compared to the
electron-electron one (since the latter conserves momentum). In particular, the
electron-phonon scattering, even if weak at $\mu=0$,  becomes   dominant at
a quite small $\mu$ such that $\mu \ll T$ [see Eq.~\eqref{small-mu}].  The temperature dependence of
conductivity away from the Dirac point is sketched in Fig.~\ref{Fig6}.
We also discuss the effect
of static disorder (assuming for definiteness that charged impurities are the
main source of disorder) as shown by dashed lines in Fig.~\ref{Fig6}. With
increasing $\mu$ (or, else, with lowering impurity concentration at fixed
nonzero $\mu$), the behavior of the Drude conductivity in the
disorder-controlled low-temperature regime changes from ``insulating''
($\sigma$ decreases at lowering $T$) to ``metallic.''

Our findings qualitatively agree with experimental data
of Refs.~\onlinecite{Bolotin} and \onlinecite{Ochoa}, see Sec.~\ref{s5} and Figs. \ref{Fig7}--\ref{Fig12}.
We hope that the results of this paper will stimulate further experimental
investigations of conductivity of suspended graphene, including a systematic
investigation of temperature dependence at different chemical potentials for $T$
up to (or even higher than) room temperature.

A number of problems related to this research have been left open.
First, this includes the possibility of essentially non-Born and
of ``ultimate quantum'' regimes at high temperatures at the Dirac point.
Second, the detailed analysis of the scattering by charged impurities with the
account of self-consistent  screening is required.
It is also interesting to study the effect of flexural phonons on quantum
transport in suspended graphene in transverse magnetic fields.

\section{Acknowledgments}
We thank A.P. Dmitriev, F. Evers, F. von Oppen,  J. Schmalian, and M. Sch\"{u}tt for useful
discussions.   The work was  supported by RFBR, by programs of the RAS, by DFG CFN, DFG SPP
``Graphene'', and by BMBF.

\appendix
\section{Calculation of scattering rate }
\label{a1}
In this Appendix, we present a derivation of the
transport time due to scattering off flexural phonons.

\subsection{Harmonic approximation}
\label{a1-1}

We first neglect the anharmonicity. The  quasistatic
random potential   representing approximately the displacement field of flexural
phonons has the form:
\be
V(\mathbf r)=\frac{g_1 T}{\varkappa S}\sum_{\mathbf q_1, \mathbf q_2}\frac{\mathbf q_1\mathbf q_2}{q_1^2 q_2^2} \sin(\mathbf q_1 \mathbf r +\f_{\mathbf q_1})\sin(\mathbf q_2 \mathbf r +\f_{\mathbf q_2}).
\label{static-pot}
\ee
[Eq.~\eqref{static-pot} is obtained by substitution of Eq.~\eqref{static-field} into Eq.~\eqref{V}.]
Averaging  squared matrix element of transition between $\psi_{\mathbf k, \alpha}$ and  $\psi_{\mathbf k', \beta}$  over  the phases $\f_\mathbf q,$ we find that  $\mathbf k^\prime= \mathbf k \pm  \mathbf q_1 \pm \mathbf q_2,$ where all four combinations of $+$ and $-$ in front of  $\mathbf q_1$ and $\mathbf q_2$ should be taken into account:
\be  \label{V-squared}
\langle \vert V_{\mathbf k,\mathbf k' }^{\alpha,\beta }\vert^2 \rangle_{\f_\mathbf q} =
\frac{g_1^2 T^2}{8\varkappa^2 S^2}\hspace{-2mm}
\sum_{ \mathbf q_1, \mathbf q_2, \pm}\hspace{-2mm} \frac{(\mathbf q_1 \mathbf q_2)^2}{q_1^4 q_2^4} \vert \langle \chi_\mathbf k ^\alpha
\vert \chi_{\mathbf k'}^\beta \rangle  \vert^2 \delta_{\mathbf k', \mathbf k \pm
 \mathbf q_1 \pm \mathbf q_2}\,.
\ee
The transport scattering rate for an electron in the branch $\alpha$ with the energy $\epsilon = \alpha v k$ is given by
\BEA
\label{tau-tr}
\frac{1}{\tau_{\rm{tr}}^\alpha(\epsilon)} &=& \frac{\pi g_1^2 T^2}{4\hbar\varkappa^2}
\sum_{ \pm,\beta}  \int \frac{d^2\mathbf q_1}{ (2\pi)^2}\frac{d^2\mathbf q_2}{
(2\pi)^2} \vert \langle \chi_\mathbf k ^\alpha \vert \chi_{\mathbf k \pm  \mathbf q_1 \pm
\mathbf q_2}^\beta \rangle  \vert^2
\nonumber \\ \nonumber
& \times &
 \frac{(\mathbf q_1 \mathbf q_2)^2}{q_1^4 q_2^4}\delta (\epsilon^\beta_{\mathbf k \pm
\mathbf q_1 \pm \mathbf q_2}-\epsilon^\alpha_{\mathbf k})(1-\mathbf n_\mathbf k
\mathbf n_{\mathbf k \pm  \mathbf q_1 \pm \mathbf q_2}). \\
\EEA
In the quasielastic  approximation,  $\alpha=\beta$ due to delta-function in
Eq.~\eqref{tau-tr}   and ${\tau_{tr}}$ is the function of $|\epsilon| =v k.$
Taking into account that the dominant contribution to the integral comes from
the region where one of the momenta is much smaller than another one, say $q_2
\ll q_1, $ and using  $ \vert \langle \chi_\mathbf k ^\alpha
\vert \chi_{\mathbf k'}^\alpha \rangle \vert^2 = (1+\mathbf n_\mathbf k \mathbf
n_{\mathbf k' })/2$, we reduce Eq.~\eqref{tau-tr} to the form
 \BEA
 \label{tau-tr-simple}
\frac{1}{\tau_{\rm{tr}}(\epsilon)} &=& \frac{\pi g_1^2 T^2}{8\hbar\varkappa^2}\sum_{ \pm,\beta}  \int \frac{d q_2}{ 2\pi q_2}
\frac{d^2\mathbf q_1}{ (2\pi)^2 q_1^2}
 \nonumber
\\
&\times &
  \delta (\epsilon^\alpha_{\mathbf k \pm  \mathbf q_1 }-\epsilon^\alpha_{\mathbf
k})[1-(\mathbf n_\mathbf k \mathbf n_{\mathbf k \pm  \mathbf q_1 })^2].
\EEA
Using  the identity \be \delta ( k - |\mathbf k+ \mathbf q|)= {\delta (q- 2k
\cos\varphi)}/ { |\cos\varphi|} \label{ident}\ee
(here $\varphi$ is the angle between $\mathbf q$ and $\mathbf k$),
we get
\BEA \label{tau-tr-q1q2}
\frac{1}{\tau_{\rm{tr}}(\epsilon)} &=& \frac{ g_1^2 T^2}{32\pi^2\hbar^2\varkappa^2 v}\int \frac{d q_2}{  q_2} \int \frac{d q_1}{  q_1}\int_{-\pi/2}^{\pi/2}
d\varphi
\nonumber
  \\
 &\times&  \frac{\delta(q_1-2k\cos\varphi)}{\cos\varphi} (1-\cos^22\varphi)\,.
\EEA
Finally, taking also into account contribution of the region $q_2\gg q_1$, we
obtain
\be
\frac{1}{\tau_{\rm{tr}}(\epsilon)}= \frac{ g_1^2 T^2}{16 \pi \hbar |\epsilon| \varkappa^2}
\ln \left(
\frac{q_{{\rm max}}}{q_{{\rm min}}} \right)\,,
\label{app-tau-tr-final}
\ee
where the energy $\epsilon$ is counted from the Dirac point, $q_{\rm{min}}\propto
1/L$ and $q_{\rm{max}} \propto |\epsilon|/v$ are infrared and ultraviolet cutoffs, respectively.
This yields Eq.~\eqref{tau-tr-final} of the main text.

\subsection{Including anharmonicity}
\label{a1-2}

Now we include the anharmonicity of flexural phonons and derive Eq.~\eqref{tau-tr-appendix} of the main text.
In order to treat both harmonic and anharmonic regions of momenta,
we introduce in Eq.~\eqref{tau-tr} a factor
$\Theta(q_1)\Theta(q_2), $ where
\be
\Theta(q)= \left \{ \begin{array}{ll}
 1, ~~\text{for} ~~ q \gg q_c   \vspace{2mm} \\ Z(q/q_c)^\eta,~~\text{for} ~~q \ll  q_c. \end{array} \right.
\ee
We also use the identity
\be 1=\int \hspace{-1mm} d^2\mathbf Q \delta (\mathbf Q \pm \mathbf q_1 \pm \mathbf q_2)= \hspace{-1mm} \int  \frac {d^2\mathbf Q d^2\mathbf r}{(2\pi)^2} \exp[i(\mathbf Q \pm \mathbf q_1 \pm \mathbf q_2)\mathbf r] ,\ee  which allows us to replace $\pm  \mathbf q_1 \pm \mathbf q_2$ with $\mathbf Q$ in  $\epsilon^\beta_{\mathbf k \pm  \mathbf q_1 \pm \mathbf q_2},~
\mathbf n_{\mathbf k \pm  \mathbf q_1 \pm \mathbf q_2}$ and $\chi_{\mathbf k \pm  \mathbf q_1 \pm \mathbf q_2},$
and  integrate in Eq.~\eqref{tau-tr} first over $d^2 \mathbf q_1 d^2 \mathbf q_2.$ Denoting the result of integration as $\xi(r)$ we find:
\BEA
&&\xi(r) \nonumber
\\&&= \sum_{\pm} \int \frac{d^2 \mathbf q_1 }{(2\pi)^2}\frac{d^2 \mathbf q_2 }{(2\pi)^2} \frac{(\mathbf q_1\mathbf q_2)^2}{q_1^4 q_2^4} \Theta(q_1) \Theta(q_2) e^{i( \pm \mathbf q_1 \pm \mathbf q_2)\mathbf r} \nonumber \\
&& =\frac{1}{2\pi^2}\left \{     \left[ \int_0^\infty \frac{dq}{q}   \Theta(q)J_0(qr)   \right]^2 \hspace{-1mm}+\hspace{-1mm}\left[ \int_0^\infty \frac{dq}{q}   \Theta(q) J_2(qr)   \right]^2\right\} \nonumber \\
&& \approx
 \frac{1}{2\pi^2}\left \{ \begin{array}{ll}
 \displaystyle{\ln^2\left (\frac{1}{q_c r} \right)}, ~~\text{for} ~~ q_c r \ll 1 \vspace{2mm} \\ \xi^* Z^2 \displaystyle{\left (\frac{1}{q_c r}\right)^{2\eta}},~~\text{for} ~~ q_c r \gg 1,
\label{approx}
\end{array} \right.
\EEA
where $\xi^* =\left[ \int_0^\infty {dx}{x^{\eta-1}}    J_0(x)   \right]^2+\left[ \int_0^\infty {dx}{x^{\eta-1}}    J_2(x)   \right]^2\simeq 2.77$ (for $\eta=0.72$).
Next, we substitute Eq.~\eqref{approx} into   Eq.~\eqref{tau-tr} and  integrate over angle of vector $\mathbf r. $ We get
\BEA  \frac{1}{\tau_{\rm{tr}}(\epsilon)}&=& \frac{g^2T^2}{\pi \hbar}\label{tau-tr_Q}
\\ &\times&  \int \frac{d^2\mathbf Q}{(2\pi)^2Q^2}   \delta (\epsilon^\alpha_{\mathbf k +  \mathbf Q }-\epsilon^\alpha_{\mathbf k})[1-(\mathbf n_\mathbf k \mathbf n_{\mathbf k +  \mathbf Q })^2] A(Q), \nonumber \EEA
where
\BEA
A(Q)&=&4\pi^2 Q^2\int d^2\mathbf r e^{-i\mathbf Q \mathbf r} \xi(r) \label{A}
\\& =&8\pi^3Q^2\int_0^\infty dr r J_0(Qr) \xi(r).
\nonumber
\EEA
By using  property of Bessel function  $xJ_0(x)=d[xJ_1(x)]/dx$ we integrate by part and find
asymptotics  of the function $A(Q)$
\BEA &&A(Q)\approx \label{approxA} \\ \nonumber
&& 8\pi\left \{ \begin{array}{ll}
 \displaystyle{\ln\left (\frac{Q}{q_c } \right)}, ~~\text{for} ~~ Q \gg q_c \vspace{2mm} \\
  \xi^* \eta Z^2 ~ \displaystyle{\left (\frac{Q}{q_c }\right)^{2\eta}}\int_0^\infty dx x^{-2\eta} J_1(x),~~\text{for} ~~ Q  \ll q_c.
\end{array} \right. \EEA

Using Eq.~\eqref{ident}
we find from Eq.~\eqref{tau-tr_Q},
\BEA
 \frac{1}{\tau_{\rm{tr}}(\epsilon)}&=&  \frac{ g^2 T^2}{4\pi^3\hbar^2 v} \nonumber \\   \hspace{-1mm} &\times&   \int \hspace{-2mm} \frac{d Q}{  Q}
 \hspace{-1mm}\int_{-\pi/2}^{\pi/2} \hspace{-3mm}d\varphi \frac{\delta(Q-2k\cos\varphi)}{\cos\varphi} (1-\cos^22\varphi)A(Q)  \nonumber \\
&=&\frac{g^2 T^2}{2\pi^3 \hbar |\epsilon|}\int_{-\pi/2}^{\pi/2} d\varphi\sin^2(\varphi) A(2k\cos\varphi).
\label{tau-tr-Q1}
\EEA
Substituting here asymptotics  of $A(Q)$ we finally obtain Eq.~\eqref{tau-tr-appendix} with the coefficient $C$ given by the following equation:
\BEA
C&=&\frac{2^{2\eta+1 }\xi^*\eta}{\pi} \int_0^\infty dx x^{-2\eta} J_1(x)\int_{-\pi/2}^{\pi/2}\hspace{-3mm}d\varphi\sin^2\hspace{-1mm}\varphi \cos^{2\eta}\hspace{-1mm}\varphi \nonumber \\ \nonumber &\simeq& 2.26.
\EEA

\subsection{Role of the screening}
\label{a1-3}

Here, we calculate the phonon-induced transport scattering rate in the presence of
the screening due to e-e interaction.
We start with the case of $\mu\ll T$.
 Replacing in Eq.~\eqref{TF4} $e^2/\hbar \kappa v$ with
renormalized value $g_e$ and using expression for $\Pi$ in the Dirac point
obtained in Ref.~\onlinecite{Schuett-ee} we conclude that screening results in
the following replacement:
\be
g \to \frac{g}{ 1+g_e N f(Q/2q_T)},
\label{TF2}
\ee
where \cite{Schuett-ee}
\BEA
  &f(z)& =  \frac{1}{\pi}\nonumber \\&\times&\hspace{-4mm}\int_1^\infty
\hspace{-3mm}du \hspace{-1mm}\int_0^1 \hspace{-3mm}dv \frac{\sinh(uz)
v(1-v^2)+\sinh(vz)u(u^2-1)}{[\cosh(uz)+\cosh(vz)]uv\sqrt{(u^2-1)(1-v^2)}}
 \nonumber \\
&=& \left\{ \begin{array}{ll} \ln2/z,~~\text{for}~~ z \ll 1   \\
\pi/8,~~\text{for}~~ z \gg 1.  \end{array} \right.
\label{f}
\EEA
The screening leads to the following
modification of the expression Eq.~\eqref{tau-tr-Q1} for the phonon-induced
scattering rate:
\be
 \frac{1}{\tau_{\rm{tr}}^{\rm{ph}}(\epsilon)}=\frac{g^2 T^2}{\pi^3 \hbar
|\epsilon|}\int_{0}^{\pi/2} d\varphi \frac{\sin^2(\varphi)
A(2k\cos\varphi)}{[1+g_eNf(\epsilon \cos\varphi/T)]^2}.
\label{tau-tr-Q2}
\ee
Combining Eqs.~\eqref{approxA}, \eqref{f}, and \eqref{tau-tr-Q2},  we obtain Eq.~\eqref{tau-tr-ph} of the main text.

Away from the Dirac point, for $\mu \gg T,$ the Thomas-Fermi screening
is accounted for by the replacement \eqref{TF1},
which can be rewritten as
\be
g \to \frac{g}{ 1+\sqrt{G_e}  k_F/\sqrt{C_3} Q }.
\label{TF1-app}
\ee
Then Eq.~\eqref{tau-tr-Q1}  becomes
\be
 \frac{1}{\tau_{\rm{tr}}^{\rm{ph}}(\epsilon)}=\frac{g^2 T^2}{\pi^3 \hbar |\epsilon|}\int_{0}^{\pi/2} d\varphi ~ \frac{\sin^2(\varphi) A(2k\cos\varphi)}{\displaystyle{\left(1+\frac{k_F\sqrt{G_e}}{2\sqrt{C_3}k\cos\varphi}\right)^2}}.
\label{tau-mu}
\ee
We see that screening can be neglected for $G_e \ll 1,$ while for $G_e \gg 1$ we get
\be
 \frac{1}{\tau_{\rm{tr}}^{\rm{ph}}(\epsilon)}=\frac{g^2 T^2 C_3|\epsilon|}{\pi^3 \hbar G_e \mu^2}\int_{0}^{\pi/2} d\varphi ~ {\sin^2(2\varphi)  A(2k\cos\varphi)}.
\label{tau-mu-1}
\ee
From Eqs.~\eqref{approxA} and \eqref{tau-mu-1} we find  that for $G_e \gg 1$   Eq.~\eqref{tau-tr-appendix}  is replaced with Eq.~\eqref{tau-tr-appendix-scr} of the main text,
where $$\tilde{C}=C C_3 \frac{\int_{-\pi/2}^{\pi/2}d\varphi\sin^2 2\varphi \cos^{2\eta}\varphi}{\int_{-\pi/2}^{\pi/2}d\varphi\sin^2\varphi \cos^{2\eta}\varphi}
\simeq 4.$$

\section{Hydrodynamic approach}
\label{a2}

Here we present details of the hydrodynamic approach used for calculation
of the conductivity in Sec.~\ref{s4} (see also
Refs.~\onlinecite{Sachdev,aip,fluid,Foster,Ryzhii, svintsov,drag,69}). We
start from kinetic equation (we keep $\hbar=1$ throughout calculations restoring
it in the final equations)
\be
\frac{\p n_\alpha}{\p t} +e\mathbf E \frac{\p n_\alpha}{\p \mathbf k} =({\hat {\text I}}_{\rm{ee}} + {\hat{\text I}}_{\rm{ph}})n_\alpha,
\label{kin}\ee
where  ${\hat {\text I}}_{\rm{ee}}$ and ${\hat {\text I}}_{\rm{ph}}$  are electron-electron and electron-phonon collisions integrals and the index $\alpha=\pm$ labels bands of positive and negative energies.
Let us introduce two  currents:
\BEA \label{Jdef}\mathbf J&=&\sum_\alpha\int d^2\mathbf k ~ n_\alpha ~\mathbf k,  \\\mathbf j&=&\sum_\alpha\int d^2\mathbf k ~ n_\alpha~\mathbf v . \label{jdef}\EEA
In a conventional semiconductor with quadratic spectrum, these currents are proportional to each other.
Importantly, this is not the case for graphene, so that velocity may relax even for a momentum conserving
scattering such as electron-electron scattering.
For simplicity, we will assume that $1/\tau_E$ due to electron-electron scattering is much larger
than other scattering rates. The peculiarity of the kinematics of particles with linear dispersion
yields fast equilibration of carriers within a given velocity direction.\cite{Kashuba,Fritz,Sachdev,aip}
Therefore, the electron gas in graphene is described  by the Fermi distribution  function
characterized by local  temperature and chemical potential both depending  on the velocity angle.
Using two variables,\cite{drag,69} electron energy and  the velocity unit vector $\mathbf{\hat{v}}=\mathbf{v}/v$,
instead of electron momentum and band index $\alpha$, the distribution function takes the form
\be
n(\epsilon,\mathbf {\hat v})=\frac{1}{\exp\{[\epsilon -\mu(\mathbf{\hat{v}})]/T(\mathbf{\hat{v}})\}+1}.
\label{n}\ee
We assume that electric field $\mathbf E$ is small, so that
$T(\mathbf{\hat{v}})-T \propto  \mathbf E\mathbf v, ~~ \mu(\mathbf{\hat{v}})-\mu \propto  \mathbf E\mathbf v .$
Expanding Eq.~\eqref{n} up to the first order with respect to $\mathbf E,$
we find the following expression for correction to the Fermi distribution function:
\be
\delta n=-\frac{\p n_F}{\p \epsilon} \frac{e\mathbf E\mathbf v}{T} \chi.
\label{dn}\ee
Here
\be
\chi=\chi_0 + \chi_1 \epsilon/T ,\label{chi}\ee
 and $\chi_0, \chi_1$ are energy-independent amplitudes (in a non-stationary case, these amplitudes  depend on time).
The currents \eqref{Jdef} and  \eqref{jdef}  may be written as
 \be \mathbf J =\langle \mathbf J_{\epsilon} \rangle, ~~\mathbf j=\langle \mathbf j_{\epsilon} \rangle, \label{Jj}\ee
 where the current densities in the energy space $\mathbf J_{\epsilon}$ and $ \mathbf j_{\epsilon}$ are
  expressed in terms of $\chi_0$ and $\chi_1:$
 \BEA  && \mathbf J_{\epsilon}=\frac{e\mathbf E T}{2}\left(  \frac{\epsilon}{T}\chi_0 +\frac{\epsilon^2}{T^2}\chi_1 \right), \label{J}\\&&  \mathbf j_{\epsilon}=\frac{e\mathbf E v^2}{2}\left(  \chi_0 +\frac{\epsilon}{T}\chi_1 \right), \label{j}\EEA
and $\langle \cdots \rangle$ is given by Eq.~\eqref{ave}.
Due to the fast energy relaxation one may  reduce the kinetic equation to the simple balance equations for $\mathbf J$ and  $\mathbf j,$  or, equivalently, to the equations for  $\chi_0$ and $\chi_1.$ To this end, we multiply Eq.~\eqref{kin} by $\mathbf k$ and $\mathbf v$ and integrate over energy and velocity angle taking into account that electron-electron collisions conserve momentum, while velocity may relax.   The result reads (see Refs.~\onlinecite{Schuett-ee,drag} for discussion of the properties of  $\hat I_{\rm{ee}}$)
\BEA  \hspace{-8mm} && \frac{\p \mathbf J}{\p t} -\frac{e\mathbf E \left \langle \epsilon \right \rangle}{2} = - \left \langle  \frac{\mathbf J_{\epsilon}}{\tau_{\rm{ph}}(\epsilon)} \right\rangle,
\label{Jdynamics}
\\
\hspace{-8mm}&& \frac{\p \mathbf j}{\p t}
-\frac{e\mathbf E v^2\hspace{-1mm}\left\langle 1 \right \rangle}{2} = -\left\langle \hspace{-1mm} \frac{\mathbf j_{\epsilon}- \epsilon~ d\mathbf j_{\epsilon}/d\epsilon}{\tau_{\rm{ee}}(\epsilon)} \hspace{-1mm}\right\rangle - \left\langle \hspace{-1mm}\frac{\mathbf j_{\epsilon}}{\tau_{\rm{ph}}(\epsilon)} \hspace{-1mm}\right\rangle. \label{jdynamics}
\EEA
By using Eqs.~\eqref{J}--\eqref{jdynamics}  one may derive equations describing  relaxation of the amplitudes $\chi_0$ and $\chi_1$ to their stationary values. The latter can be   found from the following set of equations:
\BEA     \left \langle \frac{\epsilon}{T} \right\rangle &=& \chi_0\left \langle \frac{\epsilon}{T}\frac{1}{\tau_{\rm{ph}}} \right\rangle   +\chi_1\left \langle \frac{\epsilon^2}{T^2}\frac{1}{\tau_{\rm{ph}}}  \right\rangle,
\label{Jstat}
\\
\left\langle 1 \right \rangle &=& \chi_0\left\langle \frac{1}{\tau_{\rm{ee}}} +\frac{1}{\tau_{\rm{ph}}}\right\rangle +\chi_1 \left\langle \frac{\epsilon}{T}\frac{1}{\tau_{\rm{ph}}} \right\rangle \label{jstat}.\\&&\nonumber
\EEA
The solution of Eqs.~\eqref{Jstat} and \eqref{jstat} should be substituted into  the expression for conductivity,
\be \sigma = \frac{e^2v^2N}{2} \left[\chi_0 \left\langle 1\right\rangle  + \chi_1 \left\langle \frac{\epsilon}{T}\right\rangle\right], \label{sigma-chi}\ee
which directly follows  from Eqs.~\eqref{jdef}, \eqref{dn}, and \eqref{chi}.

For $\mu=0,$ coefficients  $ \left \langle {\epsilon}/{T} \right\rangle,\left \langle {\epsilon}/{T}{\tau_{\rm{ph}}} \right\rangle$ including averaging of odd functions of energies turn to zero, so that we find \be \chi_0= \frac{\langle1\rangle}{\left\langle {1}/{\tau_{\rm{ee}}} +{1}/{\tau_{\rm{ph}}}\right\rangle} ,~~\chi_1=0,\ee
and restore Eq.~\eqref{Drude-competition0} for conductivity.

In the opposite limiting case, $\mu \gg T,$ all averages entering Eqs.~\eqref{Jstat} and \eqref{jstat} are   calculated with the use of equation $\langle A(\epsilon)\rangle \approx A(\mu)\rho(\mu)$ and we find
\be
\chi_0=0~,\chi_1 =\tau_{\rm{ph}}(\mu) \frac{T}{\mu}.\ee
The conductivity is given by
\be
\sigma=\frac{e^2N\mu\tau_{\rm{ph}}(\mu)}{4\pi\hbar^2}
\label{si}
\ee
and does not depend on the rate  of electron-electron collisions.
Using Eqs.~\eqref{tau-tr-appendix-scr} and \eqref{si} we arrive to Eq.~\eqref{sigma-mu} for the
conductivity.

Consider now the case $0<\mu \ll T$.   Remarkably, a new regime arises  in this
region provided that electron-electron collisions dominate over phonon
scattering at $\mu=0.$  To see this, we first notice that $ \langle
{\epsilon}/{T} \rangle \sim \mu/T,~~\langle {\epsilon}/{T}{\tau_{\rm{ph}}}\rangle
\sim (\mu/T) \langle {1}/{\tau_{\rm{ph}}} \rangle$ and $\langle
{\epsilon^2}/{T^2}{\tau_{\rm{ph}}}\rangle  \sim  \langle {1}/{\tau_{\rm{ph}}} \rangle,$
where  $\langle1/
{\tau_{\rm{ph}}}\rangle$  is calculated for $\mu=0$.  Then from Eqs.~\eqref{Jstat}--\eqref{sigma-chi} we find Eq.~\eqref{small-mu}.


\begin{thebibliography}{99}
\bibitem{footnote} The  effect of the virtual  flexural phonons on renormalization is governed by  the  parameter $g\Delta/\Delta_*\sim 10^{-2}$ and hence is negligibly small.
\bibitem{comment} Such a possibility is suggested by calculation of the
self-energy within the
self-consistent Born approximation (where ultraviolet divergent integrals are cut off by Fermi momentum),
which yields  both  real and imaginary part of the self-energy  scaling
with energy as $|\epsilon|^\eta. $
\end{thebibliography}
\end{document}